# WUI-CAST: Real-Time Forecast Updating for Partially Observed Wildland–Urban Interface Fires

**Can Wang[1], Naiyu Wang[1,✉], Yang Zhou[1]**

[1] College of Civil Engineering and Architecture, Zhejiang University, Hangzhou, China

✉Corresponding author: naiyuwang@zju.edu.cn

**ABSTRACT** Wildland–urban interface (WUI) fires couple landscape spread, structural fire development, and wind-driven ember transport. Forecasting building risk requires models that incorporate emerging observations and update predictions as fires evolve. This study presents WUI-CAST, a framework integrating established fire-propagation representations with sequential Bayesian state estimation to support real-time forecast updating under partial observation. A Dynamic Bayesian Network represents nonlocal ember-driven ignition and local structural propagation, while Bayesian filtering updates building-state probabilities using sparse reports. A retrospective Camp Fire application provides a proof-of-concept evaluation. Open-loop simulations captured key features of the wildfire-to-community transition. In two 30-minute forecasts evaluated against subsequently reported buildings before their assimilation, closed-loop updating increased Recall@10% from 29.03% to 83.87% and from 0.00% to 43.75%, with improved conditional spatial logarithmic scores in both windows. Regional and local diagnostics reveal how assimilation redistributes risk beyond directly observed buildings. Controlled experiments demonstrate gains across the tested observation conditions and filtering implementations. A forecast–update cycle averaged 42 s for a 5-min simulation interval. Together, these findings demonstrate improved spatial risk prioritization from sparse observations and a computationally feasible approach to real-time WUI forecast updating.

## 1 INTRODUCTION

Infrastructure systems exposed to extreme hazards exhibit cascading, time-dependent failure dynamics, in which local disruptions propagate across interacting components and escalate into system-level losses. In such settings, uncertainty does not remain static; it co-evolves with the system state as hazard intensity, component response, and network interactions unfold. The central challenge is therefore not merely forward simulation of hazard trajectories, but real-time inference of the evolving system state under partial and noisy observation. Open-loop models propagate the system state without incorporating incoming observations; an updating mechanism is needed to use this information to correct forecast drift.

Wildland–urban interface (WUI) fires represent a particularly acute manifestation of this challenge (Dossi et al., 2026; Khakzad, 2019). These events arise from interactions between continuous wildland fuels and discrete built infrastructure, producing multiscale, cross-medium propagation mechanisms that include surface spread, radiative structure-to-structure ignition, and long-range ember transport (Ding et al., 2026; Filkov et al., 2023). In recent decades, both hazard intensity and exposure have increased due to climate change, fuel accumulation, and continued development in fire-prone regions (Rao et al., 2022; Walls and Wibbenmeyer, 2025). Catastrophic events such as the 2018 Camp Fire and the 2021 Marshall Fire demonstrate how rapidly evolving WUI hazards produce spatially clustered structural failures and severe societal disruption (Buechi et al., 2021; Maranghides et al., 2021; Szasdi-Bardales et al., 2025). These characteristics impose two simultaneous requirements on modeling frameworks: capturing cross-medium propagation mechanisms and supporting real-time probabilistic state estimation under partial observation.

During active WUI incidents, operational decisions extend beyond a single deterministic projection of perimeter growth to evolving assessments of which structures face the highest near-term risk, how these risks shift over time (Masoudvaziri et al., 2023), and how limited suppression resources may be allocated within short operational windows. Real-time WUI response is inherently sequential and uncertainty-driven: UAV reconnaissance (Boroujeni et al., 2025), field reports (Maranghides et al., 2023, 2021), and remote sensing provide intermittent (Lareau et al., 2022), incomplete, and often inconsistent information, while the hazard itself continues to propagate. As a result, decision-making occurs under rapidly changing and fragmentary information. Conventional open-loop models, and many WUI-focused implementations in particular, are limited in their ability to assimilate sparse structural observations and recursively update building-level risk states once a simulation is initiated.

This disconnect between operational needs and modeling capabilities is reflected in the structure of existing approaches. A substantial body of research has advanced WUI fire modeling; yet existing approaches emphasize different aspects of the problem and remain divided across modeling paradigms. Wildland fire models, rooted in continuous fuel representations (e.g., Rothermel-based systems, BehavePlus, FlamMap, FARSITE, WRF-Fire), provide robust tools for landscape-scale spread prediction but represent the built environment in aggregated or simplified forms, limiting their ability to resolve discrete structure-to-structure ignition processes. In contrast, urban fire models developed for dense building clusters and extended with radiation and plume dynamics (Hamada, 1975; Himoto et al., 2008; Himoto and Tanaka, 2008; Lee and Davidson, 2010a, 2010b) capture building-level interactions but are not designed to represent heterogeneous landscape–structure coupling. These complementary strengths motivate integrated representations of landscape and structural fire propagation.

Recognizing this divide, recent efforts have sought to couple wildland and structural fire processes within unified modeling frameworks. However, these advances do not yet fully address the combined requirements of sequential probabilistic updating in operational settings. Long-term frameworks (Chulahwat et al., 2022; Esparza et al., 2026; Jana et al., 2025; Mahmoud and Chulahwat, 2020, 2018) provide interpretable community-scale risk metrics but rely on aggregated hazard representations and are not designed for time-resolved forecasting during active events. High-fidelity physics-based simulators (McGrattan et al., 2000; Mell et al., 2009; Morandini et al., 2013; Morvan et al., 2011) resolve vegetation–structure interactions at fine scales but are computationally prohibitive for real-time use and are typically operated in open-loop mode. More scalable quasi-physical and cellular-automata models (Jiang et al., 2021; Masoudvaziri et al., 2021; Ohgai et al., 2007; Purnomo et al., 2024; Szasdi-Bardales et al., 2025) improve efficiency but remain predominantly deterministic, with uncertainty typically approximated through external ensembles rather than embedded within system dynamics. Although data-assimilation methods have been explored for wildland fire-front forecasting, they have primarily focused on perimeter or hazard-state correction rather than recursive estimation of building-level structural states in coupled WUI cascading spread.

The methodological gap is the integration of cross-medium propagation, uncertainty evolution, and observation-driven updating within a common probabilistic state. Linking these capabilities enables sparse reports to inform both the estimated fire state and subsequent building-risk forecasts. We introduce WUI-CAST (Wildland–Urban Interface Cascading-risk Assimilation and State-space Tracking) to meet this need. A Dynamic Bayesian Network integrates established representations of nonlocal ember exposure, local structural propagation, and internal fire development. Sequential Bayesian filtering updates the inferred system state as building reports arrive, producing evolving building-state probability fields for risk prioritization.

The principal contribution is the integration of fire propagation and sequential state estimation into a framework for adaptive building-risk forecasting during partially observed WUI fires. Its value lies in converting sparse incident reports into spatially distributed forecast corrections, including at buildings not directly observed. An in-depth, retrospective Camp Fire application provides a proof-of-concept evaluation, testing whether assimilation improves identification of subsequently reported impacts relative to the same propagation model without updating. External comparisons place propagation performance in context, posterior diagnostics explain how information influences risk across the community, and controlled experiments examine sensitivity to observation conditions and filtering implementation.

## 2 METHODOLOGY

### 2.1 Overview of the WUI-CAST Modeling Framework

Building on the structural gap identified in Section 1, we formalize WUI-CAST as a partially observed stochastic state-space system that embeds coupled fire propagation mechanisms within a unified probabilistic architecture suitable for sequential state estimation. Established surface spread, radiative transfer, and ember transport physics are incorporated into a stochastic dynamical framework that supports native uncertainty propagation and recursive Bayesian updating.

Wildland–urban interface fire propagation arises from three interacting mechanisms operating across two media:

1. **Surface spread** through vegetative fuels;
2. **Short-range structure-to-structure ignition** via radiation and flame contact;
3. **Long-range stochastic ignition** via wind-transported embers.

These processes interact through cross-medium coupling: burning regions produce embers; embers deposit and trigger new ignitions; newly ignited areas become secondary sources.

The community is decomposed into two interacting layers:

- **Landscape layer** — burning and unburned vegetative cells governing surface propagation and ember generation;
- **Structural layer** — discrete building-level fire states.

The layers are coupled through an explicit ember exposure field that mediates nonlocal interactions (Figure 1).

**State–Space Representation**

Formally, we define a latent system state at time $t$,

$$X_t = (X_t^{\text{str}}, X_t^{\text{land}}) \tag{1}$$

where $X_t^{\text{str}}$ denotes structural states, including discrete fire status for each building; $X_t^{\text{land}}$ represents the landscape fuel state, including fuel properties and burning status of spatial cells. This decomposition separates medium-specific dynamics while enabling explicit cross-layer coupling.

System evolution is governed by a stochastic transition kernel $p(X_{t+1} \mid X_t)$, which encodes landscape propagation, structural ignition and progression, and ember generation, transport, and deposition. Embers generated by burning fuels and structures are transported by wind and deposited across the domain. Deposited mass modulates structural ignition probabilities and may initiate new landscape ignitions. Conditional on exogenous environmental forcing, the full latent process $\{X_t\}$ remains first-order Markov.

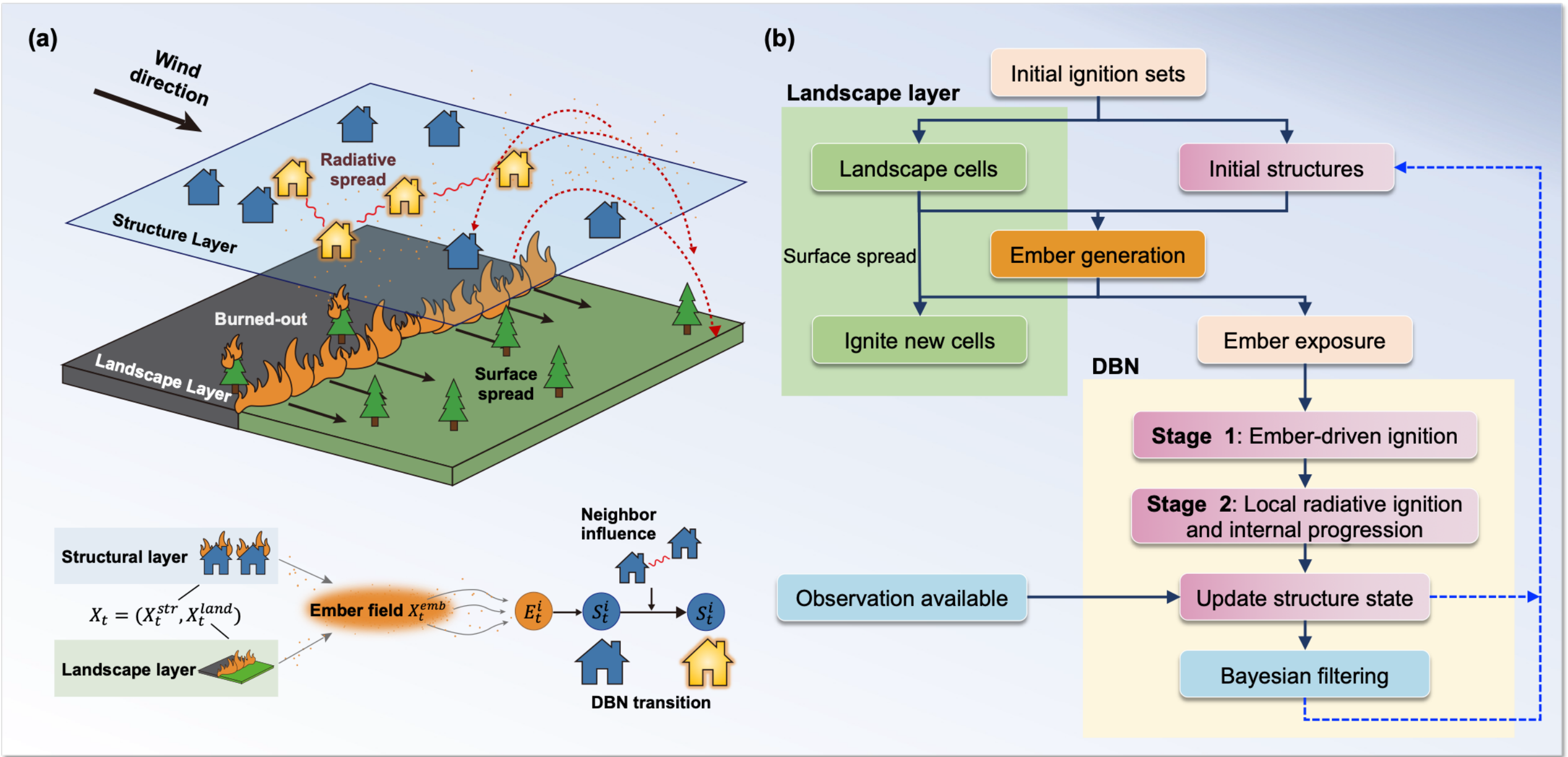

**Figure 1.** Conceptual architecture of WUI-CAST. (a) Coupled landscape–structure fire processes mediated by ember transport. (b) State–space workflow integrating prior dynamics, DBN-based structural evolution, and Bayesian filtering for closed-loop updating.

### Observation and Filtering Framework

Let $Y_t$ denote observations available at time $t$ (e.g., building damage reports or remote detections). We assume:

$$p(X_{t+1} \mid X_{0:t}, Y_{1:t}) = p(X_{t+1} \mid X_t) \quad (2)$$

$$p(Y_t \mid X_{0:t}, Y_{1:t-1}) = p(Y_t \mid X_t) \quad (3)$$

The joint distribution over $t = 1, \cdots, T$ factorizes as

$$p(X_{0:T}, Y_{1:T}) = p(X_0) \prod_{t=0}^{T-1} p(X_{t+1} \mid X_t) \prod_{t=1}^{T} p(Y_t \mid X_t) \quad (4)$$

Within this formulation, the transition kernel $p(X_{t+1} \mid X_t)$ defines the prior system evolution, the observation model $p(Y_t \mid X_t)$ defines the likelihood, and inference proceeds through recursive posterior updating via Bayesian filtering.

The primary inferential target is the filtering distribution $p(X_t \mid Y_{1:t})$, computed through the standard prediction–update recursion:

$$p(X_{t+1} \mid Y_{1:t}) = \int p(X_{t+1} \mid X_t) p(X_t \mid Y_{1:t}) dX_t \quad (5)$$

$$p(X_{t+1} \mid Y_{1:t+1}) \propto p(Y_{t+1} \mid X_{t+1}) p(X_{t+1} \mid Y_{1:t}) \quad (6)$$

### 2.2 Probabilistic Modeling of the Structural Layer

At the building scale, WUI fire propagation is governed by two physically distinct mechanisms:

1. long-range stochastic ignition driven by ember spotting;
2. short-range heat transfer between neighboring structures through radiation and flame contact.

Ember impacts are spatially nonlocal and highly variable, reflecting fluctuations in wind, source intensity, and deposition processes. In contrast, radiative and conductive heat transfer between adjacent buildings is localized and strongly conditioned on geometry, separation distance, and orientation. A probabilistic model must therefore accommodate both nonlocal forcing and local interactions.

Let $I^{\text{str}}$ denote the index set of buildings. For each building $i \in I^{\text{str}}$, define a discrete fire-state variable

$$S_t^i \in \{\text{Safe}, \text{Ignite}, \text{Fire}, \text{Burned}\}, \quad S_t = \{S_t^i\}_{i\in I^{\text{str}}} \tag{7}$$

Structural states evolve monotonically: Safe≺Ignite≺Fire≺Burned, and backward transitions are assigned zero probability. The structural layer is formulated as a Dynamic Bayesian Network (DBN) embedded within the global state-space model. The DBN encodes conditional dependencies across time and neighbors while preserving first-order Markov structure at the global state level.

### 2.2.1 Ember exposure as the nonlocal coupling input

Ember deposition provides the principal mechanism of nonlocal interaction between landscape and structures. Let $E_t^i$ denote the ember mass deposited per unit footprint area of building $i$ at time $t$.

Define $B_t = \{j \in I^{\text{str}}: S_t^j = \text{Fire}\}$ as the set of burning structures at time t, and let $C_t$ denote burning landscape cells. For each source $r \in B_t \cup C_t$, let $m_t(r)$ denote its ember production intensity. Conditional on wind and atmospheric conditions, ember transport is described by a dispersion kernel $K_t(x_i \mid x_r)$, representing the fraction of that emitted mass deposited within the footprint of building $i$. Let $A_i$ denote the footprint area of building $i$. The resulting areal ember loading of building $i$ at time $t$ is

$$E_t^i = \sum_{r\in B_t \cup C_t} \frac{m_t(r) K_t(x_i \mid x_r)}{A_i} \tag{8}$$

Stochasticity arises from the random states determining source locations and intensities. Conditional on these states and wind forcing, ember transport is deterministic.

### 2.2.2 Two-stage temporal evolution

Ember spotting produces stochastic, nonlocal ignitions driven by transported embers. In the present formulation, this distinction is reflected directly in the temporal indexing: ember deposition is assumed to occur on a timescale shorter than the model time step, so that embers generated by burning sources can be treated as an effectively instantaneous nonlocal input available at time $t$. In contrast, radiative heat transfer and internal combustion govern localized spread and within-building progression over longer effective timescales, and are therefore represented through the inter-step evolution from t to $t + 1$.

To preserve this separation, we introduce an intermediate state variable for each building. For each building $i \in I^{\text{str}}$, define an intermediate state

$$Z_t^i \in \{\text{Safe}, \text{Ignite}, \text{Fire}, \text{Burned}\}$$

which represents the intermediate state after the ember-driven stage of the model step over the interval $(t, t + 1]$ but before local neighbor-driven propagation and internal fire development. This indexing ensures causal consistency: $(S_t, E_t) \to Z_t \to S_{t+1}$. Thus, each time step is decomposed into two physically interpretable stages.

**Stage 1 — Ember-Driven Update (Nonlocal Forcing)**

Conditional on the previous structural state $S_t^i$ and ember exposure $E_t^i$, the intermediate state evolves independently across buildings:

$$p(\, Z_t \mid S_t, E_t\,) = \prod_{i\in I^{str}} p(Z_t^i \mid S_t^i, E_t^i) \tag{9}$$

This factorization reflects conditional independence of ember response across buildings given local exposure. Nonlocal interactions are already encoded in the exposure field $E_t$; no additional cross-building dependence is required at this stage.

**Stage 2 — Local Propagation and Internal Progression**

Given the intermediate state $Z_t$, the structural state at time $t + 1$ evolves through localized interactions:

$$p(\, S_{t+1} \mid Z_t\,) = \prod_{i\in I^{str}} p(S_{t+1}^i \mid Z_t^i, Z_t^{N(i)}) \tag{10}$$

where $N(i)$ denotes the set of neighboring buildings capable of influencing building $i$ through radiative heat transfer or flame contact. This stage captures: neighbor-to-neighbor radiative ignition, flame contact spread, internal fire growth and burnout dynamics. Unlike Stage 1, dependence across buildings arises through neighborhood coupling $Z_t^{N(i)}$.

### 2.2.3 Structural transition kernel

The augmented one-step transition over the pair $(\,S_{t+1}, Z_t)$ factorizes as

$$p(\, S_{t+1}, Z_t \mid S_t, E_t\,) = p(\, Z_t \mid S_t, E_t\,)\, p(S_{t+1} \mid Z_t) \tag{11}$$

The structural transition kernel entering the global state-space model is obtained by marginalizing over the intermediate state:

$$p(\, S_{t+1} \mid S_t, E_t\,) = \sum_{Z_t} p(\, Z_t \mid S_t, E_t\,)\, p(S_{t+1} \mid Z_t) \tag{12}$$

Substituting the factorizations yields

$$p(\, S_{t+1} \mid S_t, E_t\,) = \sum_{Z_t} \prod_{i\in I^{str}} p(Z_t^i \mid S_t^i, E_t^i) \prod_{i\in I^{str}} p(S_{t+1}^i \mid Z_t^i, Z_t^{N(i)}) \tag{13}$$

This defines a locally interacting Markov system in which nonlocal ember forcing enters through $E_t$ (Stage 1) and neighborhood coupling enters through $N(i)$ (Stage 2), and, conditional on the exposure input, temporal evolution remains first-order Markov in $S_t$.

The two-stage construction provides physical interpretability, separates nonlocal forcing from local dependence, and improves computational tractability. Embedding this mechanism-resolved temporal structure within the Dynamic Bayesian Network preserves the causal structure of fire physics while enabling scalable probabilistic state estimation. Figure 2 illustrates the within-step factorization and the corresponding graphical structure.

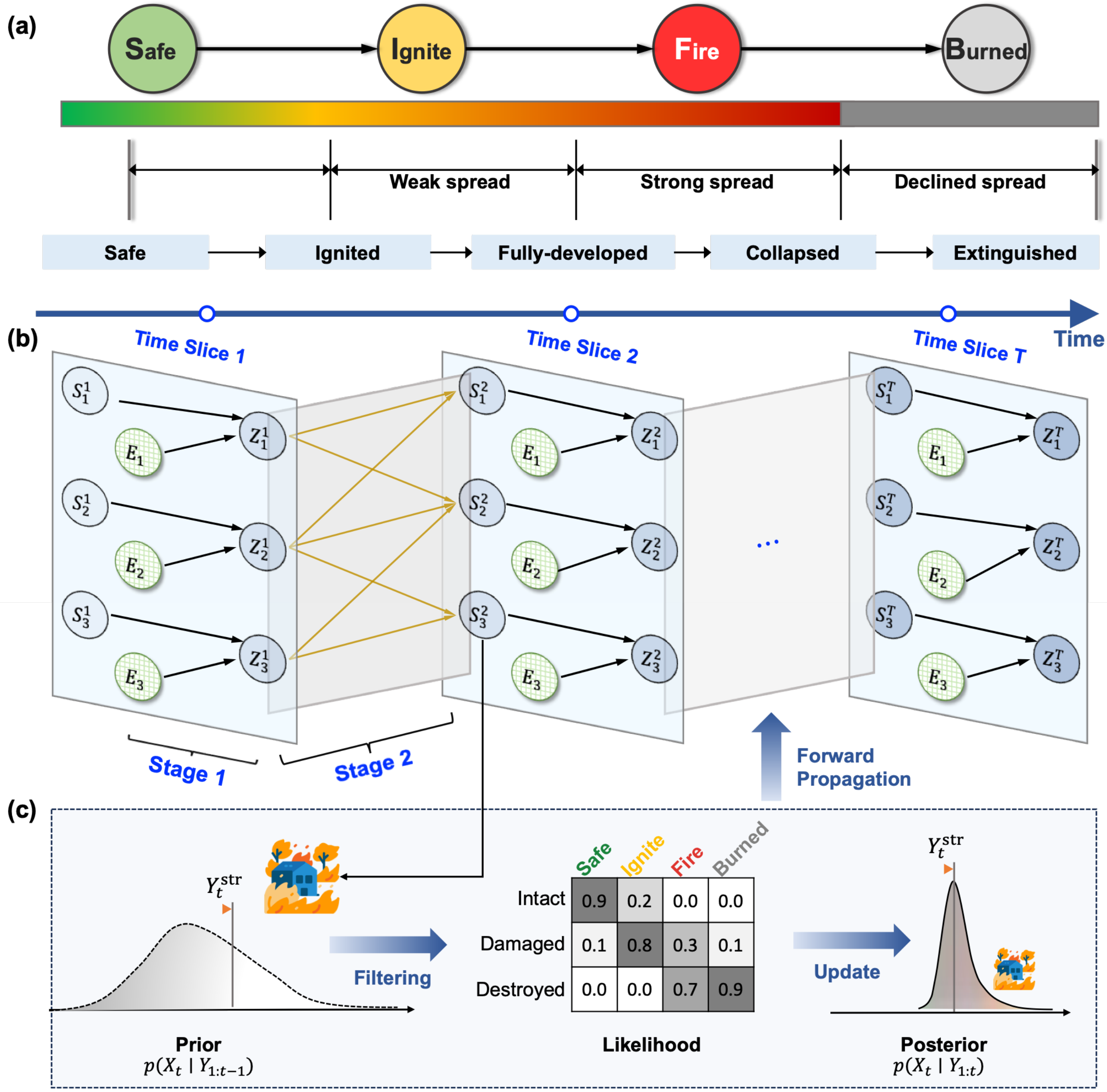

**Figure 2.** Structural Dynamic Bayesian Network for WUI-CAST. (a) Irreversible four-state building progression. (b) Two-stage state update: ember-driven ignition followed by local propagation and internal progression with neighborhood coupling. (c) Posterior building states updated via Bayesian filtering under partial observations.

#### 2.2.4 Observation model and sequential data assimilation

WUI-CAST assimilates partial and noisy observations to recursively update the posterior distribution over the latent state. Let $Y_t$ denote observations available at time $t$. Let $O_t^{\text{str}} \subset I^{\text{str}}$ denote the subset of buildings with valid observations at time $t$. In this study we focus on building-level status labels (e.g., intact, damaged, destroyed). Landscape observations could be incorporated analogously but are not used here.

For each observed $i \in O_t^{\text{str}}$, $y_t^i$ represents a case-specific structural condition report. In the Camp Fire implementation, these reports are mapped to the binary affected / unaffected reporting interface described in Section 3.4. The latent structural state is $S_t^i \in \{\text{Safe}, \text{Ignite}, \text{Fire}, \text{Burned}\}$. Because the observed categories are typically coarser than the internal four-state representation, the observation model maps latent states to observed labels through a categorical likelihood defined by a confusion matrix, which captures classification uncertainty and reporting noise.

Assuming conditional independence of observations given the latent state, the likelihood factorizes as

$$p(Y_t^{\mathrm{str}} \mid X_t) = \prod_{i \in O_t^{\mathrm{str}}} p(y_t^i \mid S_t^i) \quad (14)$$

Since only structural states are observed in this implementation:

$$p(Y_t \mid X_t) = p(Y_t^{\mathrm{str}} \mid X_t) \quad (15)$$

Buildings without observations contribute no direct likelihood factor. Their posterior probabilities may nevertheless change indirectly because observations reweight the distribution over system trajectories, and the corrected trajectories then propagate through the transition kernel via ember-driven exposure and neighborhood coupling. Thus, partial observations can reshape the posterior state field beyond the directly observed buildings.

At observation times, the likelihood in Eq. (14) enters the Bayesian filtering update defined in Section 2.1. Between observation times, the state distribution evolves under the transition kernel alone. The case-specific observation labels and reference likelihood matrix used for the Camp Fire implementation are specified in Section 3.4.

### 2.3 Physics-informed Parameterization of State Transitions

Section 2.2 established structural evolution as a two-stage stochastic process separating nonlocal ember-driven ignition from local propagation and internal fire development. We now specify physically interpretable probabilistic response functions for the two-stage structural transition process.

Rather than imposing deterministic ignition thresholds, we adopt probabilistic response functions to represent variability arising from material heterogeneity, construction differences, wind fluctuations, and unresolved sub-grid processes. This aligns with a reliability perspective, in which ignition is treated as a stochastic event conditioned on measurable physical exposure.

Throughout, transition probabilities are modeled as monotonic functions of physically meaningful inputs (e.g., ember mass, radiant heat flux, burn duration), preserving interpretability while enabling uncertainty propagation.We parameterize:

$$p\left(Z_t^i \mid S_t^i, E_t^i\right) \quad \text{(Stage 1)}$$

and

$$p(S_{t+1}^i \mid Z_t^i, Z_t^{N(i)}) \quad \text{(Stage 2)}$$

#### 2.3.1 Stage 1 — ember-driven ignition

For a generic exposure variable $x$ (e.g., ember mass or radiant energy), ignition probability is modeled using a logistic response:

$$P(\text{ignite} \mid x) = \frac{1}{1 + \exp(-k(x - x_{50}))} \quad (16)$$

where $x_{50}$ is the median ignition threshold and $k$ controls steepness.

Using $E_t^i$, the areal ember loading defined in Eq. (8), we define the ember-driven ignition probability as:

$$P_{\mathrm{emb}}^i = \frac{1}{1 + \exp(-k_{\mathrm{emb}}(E_t^i - m_{50}))} \quad (17)$$

For a safe building,

$$p(Z_t^i = \text{Ignite} \mid S_t^i = \text{Safe}, E_t^i) = P_{\mathrm{emb}}^i \quad (18)$$

$$p(Z_t^i = \text{Safe} \mid S_t^i = \text{Safe}, E_t^i) = 1 - P_{\mathrm{emb}}^i \quad (19)$$

For buildings already ignited or burning:

$$p(Z_t^i = S_t^i \mid S_t^i \in \{\text{Ignite}, \text{Fire}, \text{Burned}\}) = 1 \quad (20)$$

This enforces irreversibility at Stage 1: ember exposure cannot revert damage.

#### 2.3.2 Stage 2 — local radiative ignition and internal progression

**Radiative ignition.** If $Z_t^i = \text{Safe}$, the building $i$ receives radiative energy from burning neighbors. Let $Q_{ij}$ denote the maximum incident radiative heat-flux density across the façades of building $i$ due to burning neighbor $j$, calculated from building geometry and radiative view factors:

$$P_{\mathrm{rad}}^{(j \to i)} = \frac{1}{1 + \exp(-k_{\mathrm{rad}}(Q_{ij} - Q_{50}))} \quad (21)$$

Only neighbors in the Fire state are included in the radiative-ignition calculation. Assuming conditional independence of radiative ignition attempts given intermediate neighbor states, the total radiative ignition probability is

$$P_{\mathrm{rad,tot}}^i = 1 - \prod_{j \in N_F(i,t)} (1 - P_{\mathrm{rad}}^{(j \to i)}) \quad (22)$$

where $N_F(i,t)$ denotes the set of neighbors of building $i$ that are in the Fire state at time $t$. Thus,

$$p(S_{t+1}^i = \text{Ignite} \mid Z_t^i = \text{Safe}, Z_t^{N(i)}) = P_{\mathrm{rad,tot}}^i \quad (23)$$

$$p\left( S_{t+1}^i = \text{Safe} \,\middle|\, Z_t^i = \text{Safe}, Z_t^{N(i)} \right) = 1 - P_{\mathrm{rad,tot}}^i \quad (24)$$

**Internal fire development and burnout.** For $Z_t^i \in \{\text{Ignite,Fire}\}$, internal evolution is modeled as a continuous-time Markov process discretized over $\Delta t$. Let $\lambda_{\mathrm{grow}}$ be the growth rate (Ignite → Fire) and $\lambda_{\mathrm{ext}}$ the extinction/burnout rate (Fire → Burned). Then

$$p(Fire|Ignite) = 1 - \exp(-\lambda_{grow}\Delta t). \quad (25)$$

$$p(Ignite|Ignite) = \exp(-\lambda_{grow}\Delta t). \quad (26)$$

$$p(Burned|Fire) = 1 - \exp(-\lambda_{ext}\Delta t). \quad (27)$$

$$p(Fire|Fire) = \exp(-\lambda_{ext}\Delta t) \quad (28)$$

Burned is absorbing:

$$p(Burned|Burned) = 1 \quad (29)$$

Structural states evolve monotonically; backward transitions are assigned zero probability, as discussed in Section 2.2. Each conditional distribution is explicitly normalized. Together, these response functions complete the structural transition specification used in the state-space filtering framework.

### 2.4 Forecast Outputs and Interpretation

WUI-CAST outputs a probability distribution over the four fire states for each building, describing its most likely state and associated uncertainty. These evolving probability fields support adaptive risk ranking as new observations arrive. The case study evaluates both the propagation of these fields and the effect of sequential updating on building-risk prioritization.

## 3 CASE STUDY AND MODEL INSTANTIATION FRAMEWORK

### 3.1 Case Characteristics and Simulation Scope

The 2018 Camp Fire is selected as the case study because it represents a rapidly evolving wildland–urban interface (WUI) event in which landscape spread, ember-driven ignition, and structure-to-structure transmission interact in a tightly coupled manner (Szasdi-Bardales et al., 2024). These characteristics align directly with the multi-mechanism state-space formulation of WUI-CAST defined in Section 2.

Ignition occurred at approximately 06:30 PST on 8 November 2018 near Camp Creek Road in the Feather River Canyon (Figure 3). Under strong downslope winds and critically dry fuels, the fire advanced rapidly down-canyon toward Paradise along a well-defined topographic pathway. Within a short time, the event transitioned from wildland propagation to community-scale impact, with upstream spread conditioning subsequent urban losses.

At the community scale, fire development in Paradise was characterized by early ember-driven intrusion ahead of the main front, followed by distributed ignitions and later structure-dominated cascading spread. Buildings are interwoven with vegetation in a WUI intermix configuration, allowing nonlocal ignition and local structural interaction to operate concurrently. This progression—from upstream spread to dispersed ignition and cascading structural involvement—activates the key mechanisms represented in the transition kernel of the state-space system.

The simulation window spans 06:30–12:00 PST, covering ignition, wildfire approach, and the development of structural involvement in Paradise. By approximately 09:30 PST, the main fire front reached the eastern side of Paradise near Pentz Road, following the earlier onset of ember-driven ignitions within the community (Maranghides et al., 2021). The front continued westward and had largely passed through Paradise by approximately 14:00 PST, while residual burning and final structural damage continued afterward.

Section 4 evaluates propagation performance and the contribution of sparse observations to state estimation and subsequent forecasts during this wildfire-to-community progression.

### 3.2 Case-Specific State Representation and Domain Setup

Figure 4 summarizes the Camp Fire data inputs and their roles in the model. LANDFIRE fuel data (Rollins, 2009) define the landscape state $X_t^{\text{land}}$; Microsoft building footprints (Microsoft, 2019) define the structural state $X_t^{\text{str}}$ through individual DBN nodes; and RAWS wind observations (MesoWest, n.d.) provide the forcing governing ember transport and building-level exposure $E_t^{\text{i}}$. This mapping preserves the state definition and decomposition introduced in Section 2.1. The transition-kernel instantiation and the observation model used for updating are detailed in Sections 3.3 and 3.4, respectively.

WUI-CAST is implemented over an integrated wildland–urban domain extending from the Camp Creek Road ignition point through the Feather River Canyon to Paradise and adjacent areas. The computational domain spans approximately 1,200 km² (39.88°–39.61°N; −121.80°–−121.33°W), preserving the wildland-to-community progression required by the coupled state-space formulation. Spatial discretization is 30 m (consistent with LANDFIRE), yielding approximately 1.33 million landscape cells.

At the initial time t = 0, corresponding to 06:30 PST, the latent state $X_0 = (X_0^{\text{str}}, X_0^{\text{land}})$ is initialized as follows: all burnable landscape cells are unburned except for cells surrounding the documented ignition coordinate (−121.434, 39.815), which are initialized as burning; all buildings are initialized as $S_0^i$=Safe; and the ember exposure field is initialized to zero. Particles in the particle filter (PF), implemented using Sequential Monte Carlo (SMC), are initialized identically at $t = 0$, with stochastic divergence emerging from the probabilistic transition dynamics. This initialization defines $p(X_0)$ required by the Bayesian filtering recursion in Section 2.1.

At each time step, the landscape state is first propagated forward, followed by ember generation and transport to obtain building-level exposure. Stage 1 of the structural DBN applies this exposure, and Stage 2 resolves local interactions and internal progression. Available observations then enter the filtering update.

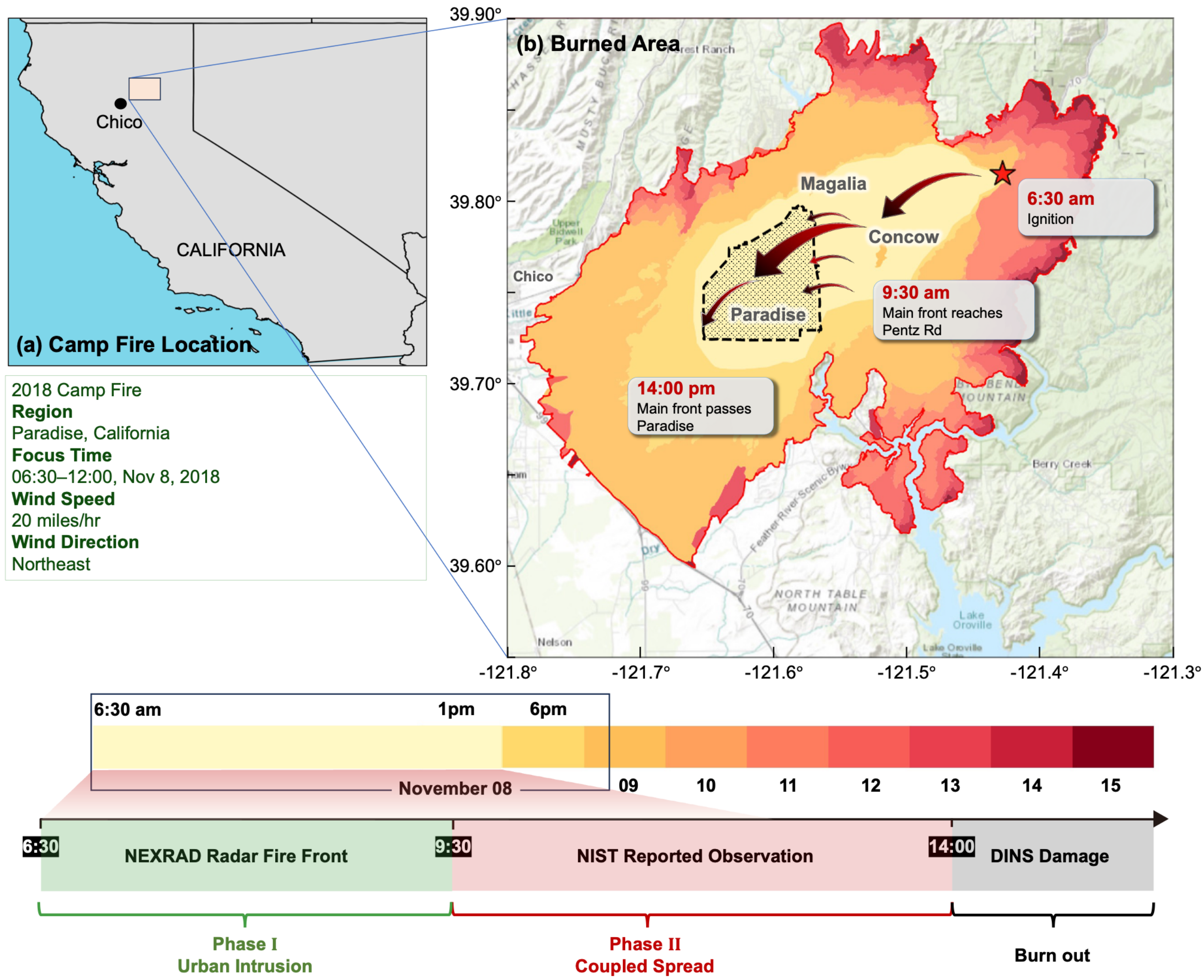

**Figure 3.** Overview and early progression timeline of the 2018 Camp Fire.

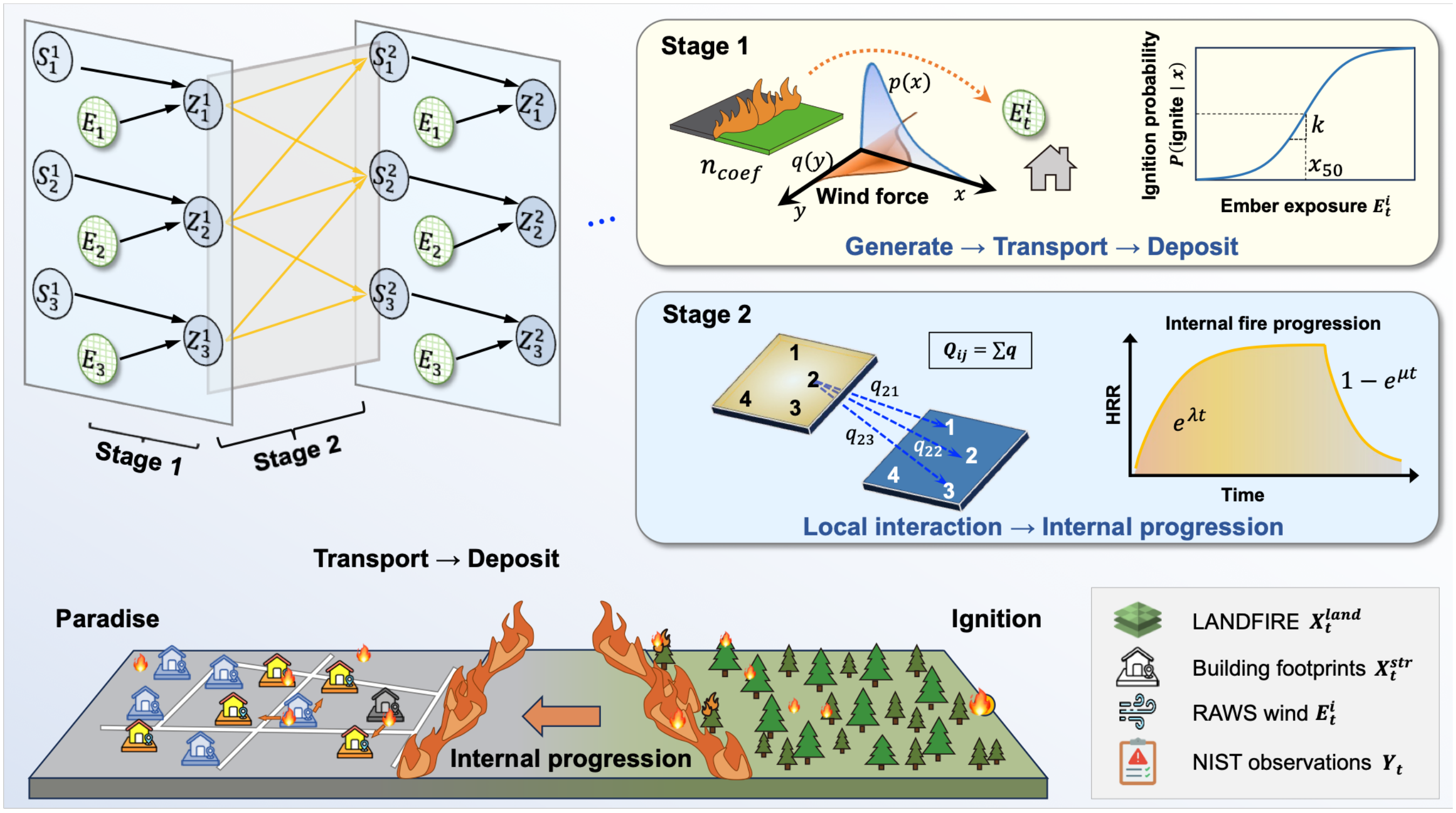

**Figure 4.** Camp Fire case instantiation of the WUI-CAST framework.

### 3.3 Transition Kernel for Probabilistic State Propagation

The Camp Fire implementation applies the transition kernel $p(X_{t+1}|X_t)$ defined in Section 2, with landscape fire spread propagated on the 30-m LANDFIRE grid using a cellular-automata formulation and structural-state transitions represented by the two-stage DBN.

Transition parameters are adopted from prior empirical or experimental studies and held fixed across scenarios, rather than calibrated to Camp Fire outcomes. RAWS wind observations prescribe the forcing for ember transport at each time step. Buildings are treated as a homogeneous class, with stochasticity represented through transition probabilities. Suppression actions and dynamic fuel-moisture effects are not explicitly represented.

Table 1 lists the simulation settings and transition parameters. Observation-likelihood parameters are specified separately in Section 3.4.

**Table 1.** Numerical instantiation of simulation setup and transition kernel parameters for the Camp Fire case.

| State–Space Component | Term (Sec. 2) | Symbol | Value | Source |
|---|---|---|---|---|
| **Domain & Discretization** | Spatial domain | — | 121.80° – 121.33°W ; 39.88° – 39.61°N | — |
| | Simulation window | $[T_{start}, T_{end}]$ | 06:30 – 12:00 PST | — |
| | Time step | $\Delta t$ | 5 min | — |
| | Ignition location | — | (121.434°W, 39.815°N) | Maranghides et al., 2021 |
| **Ember Generation (Landscape / Structure)** | Mean ember mass | $\bar{m}$ | 0.1 g | Hillen et al., 2015 |
| | Ember number coefficient | $n_{coef}$ | $4923/m^2$ | Masoudvaziri et al., 2021 |
| **Ember Transport Kernel** | Downwind displacement | $p(x)$ | $p(x) \sim LogNormal(\mu = \ln(30V), \sigma_x)$ | Masoudvaziri et al., 2021 |
| | Crosswind displacement | $q(y)$ | $q(y) \sim \mathcal{N}(0, \sigma_y)$ | |
| | Longitudinal dispersion | $\sigma_x$ | 0.3 | |
| | Lateral dispersion | $\sigma_y$ | 8 | |
| **Stage-1 : Structural Ignition** | Logistic midpoint | $m_{50}$ | $40\ g/m^2$ | Bernard et al., 2010 |
| | | $Q_{50}$ | $19\ kW/m^2$ | Lin et al., 2019 |
| | Logistic slope | $k_{emb}$ | 0.3 | Bernard et al., 2010 |
| **Stage-2 : Internal Progression** | Ignite → Fire rate | $\lambda_{grow}$ | $1/60\ min^{-1}$ | Zhao, 2011 |
| | Fire → Burned rate | $\lambda_{ext}$ | $1/180\ min^{-1}$ | |

### 3.4 Observation Model and Partial-Observation Assimilation

This subsection specifies the measurement likelihood used to assimilate partial building-level observations. As defined in the state-space formulation in Section 2.2.4, the structural filtering update requires

$$p(Y_t^{\text{str}} \mid X_t) = \prod_{i \in O_t^{\text{str}}} p(y_t^i \mid S_t^i)$$

where $O_t^{\text{str}}$ is the set of buildings observed at time $t$, $S_t^i \in \{\text{Safe}, \text{Ignite}, \text{Fire}, \text{Burned}\}$ denotes the latent state of building $i$, and $y_t^i$ is the corresponding observed condition label.

In principle, $p(y_t^i \mid S_t^i)$ should be calibrated from empirical class-conditional reporting errors. For building-scale wildfire impacts, however, no standardized observation model or benchmark dataset is currently available to estimate a reliable confusion matrix between reported conditions and latent fire-progression states. We therefore specify a weakly informative reference likelihood that provides a transparent measurement interface for recursive inference, rather than a validated empirical model of the reporting process.

In this implementation, the assimilated structural observations are mapped to a binary reporting interface: affected report versus unaffected / pseudo-safe report. The reference likelihood is defined in Table 2. The reference likelihood assigns greater affected-report probability to Fire and Burned than to Safe and Ignite, while retaining nonzero reporting uncertainty in every state. In the main NIST-assimilation experiments, only observed affected reports contribute likelihood factors. The unaffected / pseudo-safe row is retained for likelihood completeness and for the controlled pseudo-observation experiments in Section 4.4.

The NIST building-level records primarily provide positive evidence of structural impact. They do not constitute a complete inventory of unaffected buildings at each time step. Consequently, the observation process is asymmetric: a building appearing in the NIST record contributes an affected-report likelihood, whereas absence from the record is not interpreted as evidence of safety. Buildings without matched observations therefore contribute no direct likelihood factor.

These likelihoods enter the recursive update defined in Section 2.2.4. Section 4.4 evaluates sensitivity to likelihood sharpness and affected-reporting tendency.

**Table 2.** Reference observation likelihood matrix $p(y_t^i \mid S_t^i)$.

| Observed label $y_t^i$ | $S_t^i$ = Safe | $S_t^i$ = Ignite | $S_t^i$ = Fire | $S_t^i$ = Burned |
|---|---|---|---|---|
| Affected report | 0.20 | 0.30 | 0.90 | 0.80 |
| Unaffected / pseudo-safe report | 0.80 | 0.70 | 0.10 | 0.20 |

### 3.5 Experimental Design and Evaluation Framework

The experimental design separates propagation performance from the contribution of sequential updating. Open-loop WUI-CAST assesses the transition-kernel dynamics without assimilating building-level observations, whereas closed-loop WUI-CAST assimilates time-stamped observations sequentially through Bayesian filtering. FARSITE is included as an external perimeter-scale comparator over the same event window and is not intended to represent a calibrated building-level structural-impact model.

A published external WUI-model reference was also included to provide an external comparison with an existing WUI fire-spread model. We selected SWUIFT (Masoudvaziri et al., 2021) because it is a published, spatially explicit WUI fire-spread model that has been applied to the 2018 Camp Fire (Szasdi-Bardales et al., 2024) and subsequently used in other major WUI fire events, including the Marshall and Lahaina fires (Juliano et al., 2024; Szasdi-Bardales et al., 2025).

The evaluation uses three complementary data sources. NEXRAD observations (Lareau et al., 2022) provide a reference for large-scale fire-front evolution and are used to assess perimeter-scale propagation during the wildfire-approach phase. NIST building-level records provide sparse, time-stamped affected-building reports and serve as the primary input for sequential assimilation and short-horizon comparison. DINS endpoint damage data provide the final spatial distribution of burned structures and are used as an endpoint reference and for controlled pseudo-observation experiments in the robustness analysis.

The fire progression provides physical context for the evaluation rather than a strict division between nonoverlapping experimental periods. Before approximately 09:30 PST, the main front approached Paradise while ember intrusion and reported building impacts were already occurring within the community. After the main front reached the eastern community boundary near Pentz Road, landscape and structural propagation continued to interact. Accordingly, the perimeter-scale assessment spans 08:00–12:00 PST, and the community-scale analyses include observations and forecasts beginning at 08:30 PST. The two 30-min forecast tests cover 08:30–09:00 and 09:00–09:30 PST; each is scored before the report batch at its evaluation time is assimilated.

The following analyses distinguish propagation performance, the contribution of assimilation, and sensitivity to the observation conditions.

# 4 RESULTS

The Camp Fire proof-of-concept evaluation addresses four questions. Section 4.1 evaluates open-loop propagation using perimeter-scale observations and external model comparisons. Section 4.2 measures assimilation-related improvements through paired WUI-CAST forecasts, separating subsequent-report evaluation from agreement with cumulative reports. Section 4.3 examines how updates redistribute risk beyond directly observed buildings. Section 4.4 tests sensitivity to observation conditions and filtering implementation.

### 4.1 Open-Loop Propagation and External Benchmarks

We first examine WUI-CAST propagation without assimilating building reports. NEXRAD-inferred fire extent and FARSITE provide perimeter-scale references, while the published SWUIFT-USFS case provides a community-scale reference for spatial agreement. These external comparisons provide context for the paired assimilation experiments in Section 4.2.

#### 4.1.1 Perimeter-scale propagation and entry into the community

Figure 5(a) compares open-loop WUI-CAST and FARSITE perimeter evolution with NEXRAD observations from 08:00 to 12:00 PST. The FARSITE configuration follows the Camp Fire setup reported by Shamsaei et al. (2023), using LANDFIRE landscape data, Scott and Burgan fuel categories, hourly ERA5-based weather forcing, 30 m horizontal resolution, and WindNinja-derived spatially varying wind fields. IoU quantifies spatial overlap with the reference extent, whereas normalized Hausdorff distance quantifies boundary displacement; higher IoU and lower normalized Hausdorff distance indicate closer agreement.

During wildfire approach, WUI-CAST reproduces the dominant northeast-to-southwest propagation direction and elongated fire-front geometry. IoU reaches approximately 0.80 around 08:40 PST, while FARSITE shows nearly no spatial overlap with the NEXRAD-inferred extent during this period. Across the 25 evaluation times from 08:00 to 12:00 PST, WUI-CAST also has a smaller normalized Hausdorff distance than FARSITE (Figure 5(c)).

Near the transition at Pentz Road around 09:30 PST, WUI-CAST preserves a continuous spread pattern along the eastern community boundary and into Paradise. Relative area differences from the NEXRAD-inferred extent are approximately −0.55% at 09:30, −4.9% at 10:00, and −0.7% at 11:00 PST; the FARSITE extent is more than 90% smaller than the reference at those times. By 12:00 PST, WUI-CAST overpredicts the reference extent by approximately 27%. These comparisons characterize perimeter-scale agreement under the respective model configurations.

Figure 5(b) shows how the open-loop trajectory extends into the structural layer. Affected building states emerge along the eastern edge of Paradise and progress westward and inward, with dispersed involvement ahead of broader community spread. By 12:00 PST, more buildings occupy later damage states. This modeled evolution illustrates the coupling of landscape spread, ember exposure, and building-state transitions. The community-scale comparison below evaluates the spatial organization of those building-level outputs separately from the perimeter assessment.

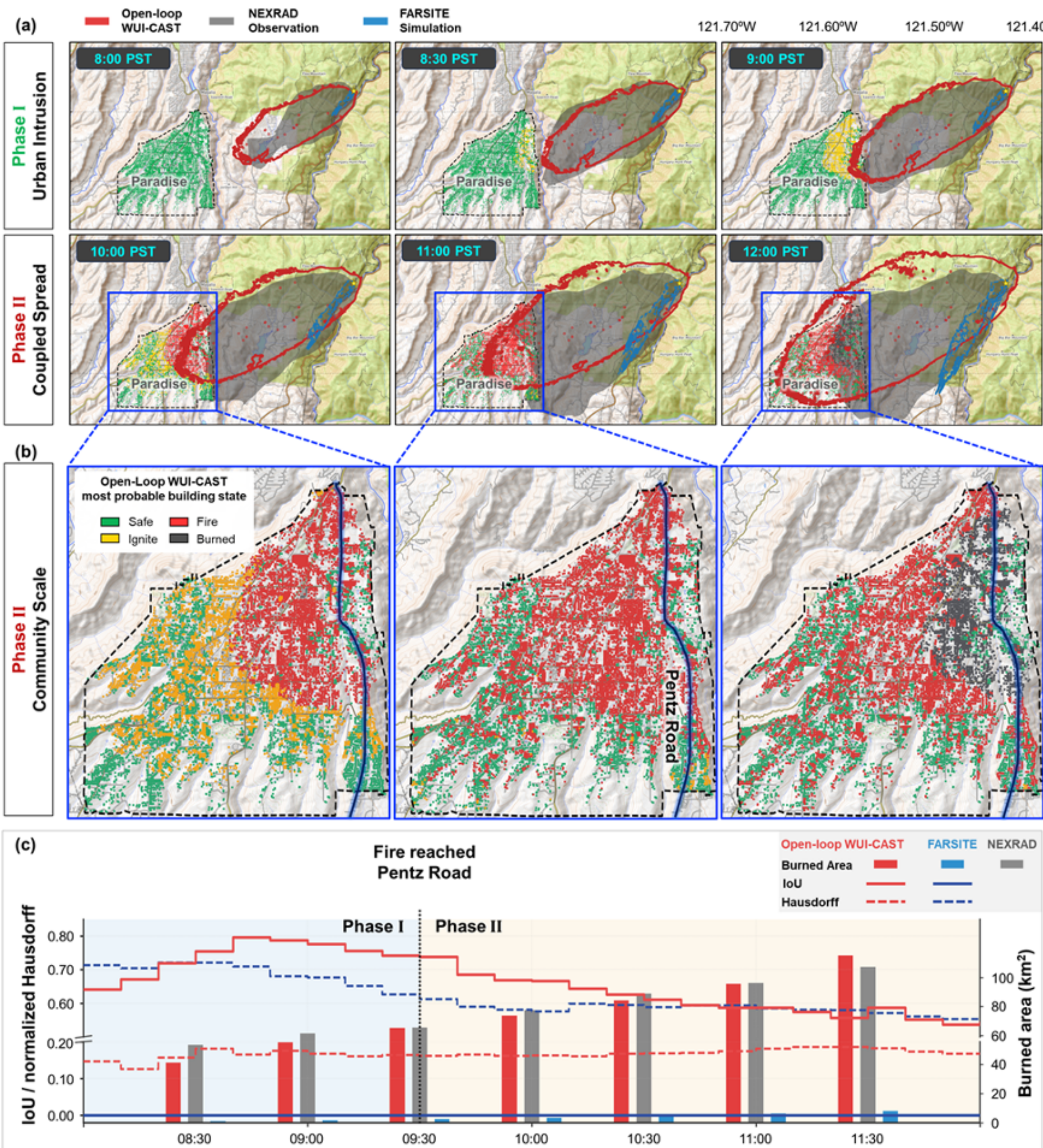

**Figure 5.** Open-loop propagation across the two phases of the Camp Fire. (a) Fire-perimeter evolution from 08:00 to 12:00 PST for WUI-CAST and FARSITE, compared with NEXRAD observations. (b) Most-probable building states in Paradise during community involvement. (c) IoU, normalized Hausdorff distance, and fire-affected area over time. FARSITE is used as a perimeter-scale comparator; its results do not evaluate building-level structural impacts.

#### 4.1.2 Community-scale comparison with the published SWUIFT benchmark

The community comparison uses the SWUIFT-USFS case reported by Szasdi-Bardales et al. (2024). Its community boundaries correspond to those adopted in WUI-CAST. WUI-CAST follows the documented initial fire conditions and other relevant settings wherever applicable to the different model formulations. The SWUIFT IoU and Sørensen-Dice values are taken directly from Table 4 of the published study; SWUIFT was not rerun or rescored. WUI-CAST values are calculated here at the same four times: 08:30, 09:00, 10:00, and 11:00 PST.

For WUI-CAST, community-scale risk is constructed from building-level affected-state probabilities. A building is classified as affected in the probabilistic representation when it occupies any of the Ignite, Fire, or Burned states:

$$p_i^{aff}(t) = P(S_t^i = Ignite) + P(S_t^i = Fire) + P(S_t^i = Burned).$$

The hotspot intensity of block b is the mean affected-state probability over its $N_b$ buildings:

$$P_b(t) = \frac{1}{N_b} \sum_{i \in b} p_i^{aff}(t).$$

A block is included in the elevated-risk extent when its intensity exceeds 1.2 times the contemporaneous mean intensity across all community blocks:

$$P_b(t) > 1.2\, P(t).$$

The multiplier of 1.2 is held constant across the four evaluation times. The selected block polygons form the WUI-CAST elevated-risk extent, which is compared with the cumulative NIST-derived observation proxy using IoU and Dice. These are related measures of spatial overlap. Matching community boundaries, evaluation times, and metric definitions supports comparison with the published SWUIFT scores, while differences in model inputs and the construction of spatial extents remain.

**Table 3.** Community-scale spatial agreement of open-loop WUI-CAST and the published SWUIFT-USFS case.

| Model | Metric | 08:30 | 09:00 | 10:00 | 11:00 | Mean |
|---|---|---|---|---|---|---|
| WUI-CAST open-loop | IoU ↑ | 0.441 | 0.652 | 0.605 | 0.523 | 0.555 |
| SWUIFT-USFS case | IoU ↑ | 0.550 | 0.560 | 0.510 | 0.490 | 0.530 |
| WUI-CAST open-loop | Dice ↑ | 0.613 | 0.790 | 0.754 | 0.687 | 0.711 |
| SWUIFT-USFS case | Dice ↑ | 0.710 | 0.720 | 0.670 | 0.660 | 0.690 |

All times are PST; ↑ indicates higher agreement. SWUIFT values, including the reported means, are quoted from Szasdi-Bardales et al. (2024), Table 4, USFS case. WUI-CAST values are calculated in this study from block-aggregated elevated-risk extents against the cumulative NIST-derived proxy.

Mean IoU and Dice are 0.555 and 0.711 for open-loop WUI-CAST, compared with the published SWUIFT values of 0.530 and 0.690. SWUIFT has higher agreement at 08:30, whereas WUI-CAST has higher agreement at the other three evaluation times. This comparison shows that open-loop WUI-CAST achieves community-scale spatial agreement comparable to the published SWUIFT case, complementing the perimeter-scale results in Figure 5. The next section holds the WUI-CAST propagation formulation fixed to examine the incremental effect of assimilating building reports.

### 4.2 Effects of Sequential Assimilation on State Estimation and Short-Horizon Forecasting

The open-loop and closed-loop configurations share initial conditions, propagation dynamics, and transition parameters, and differ in whether building reports are assimilated. This paired design addresses the contribution of sequential updating within a common model. We distinguish forecasts scored against subsequent reports from agreement with accumulated reports: the former tests whether an update helps identify buildings not yet reported at forecast issuance, whereas the latter measures how the evolving risk field represents the information accumulated during the event.

#### 4.2.1 Evaluation protocol and forecasts against subsequent reports

At each evaluation time, we rank the same fixed candidate set $\mathcal{B}$, comprising all 11,854 modeled buildings in Paradise and the immediately surrounding simulation area. Recall@10% is the percentage of reported affected buildings that fall within the highest-risk 10% of this set. It evaluates a limited-capacity prioritization task and depends on probability ordering. We additionally use a conditional spatial logarithmic score (CLS) (Gneiting and Raftery, 2007) to assess numerical spatial probability allocation. NIST reports provide positive evidence of impact but do not supply time-matched confirmed-unaffected outcomes. We therefore normalize the affected-state probabilities over the same candidate set without assigning negative outcomes to unreported buildings. For a forecast issued at time τ and evaluated at time t,

$$q_i(t \mid \tau) = \frac{p_i^{aff}(t \mid \tau)}{\sum_{j \in \mathcal{B}} p_j^{aff}(t \mid \tau)}.$$

Let $N_{\tau,t}$ denote the unique buildings first reported in the interval (τ,t], excluding buildings reported by the issuance time,

and let $H_{\tau,t}$ be the highest-risk 10% of candidate buildings in the forecast. The two scores for each model configuration are

$$Recall@10\%(\tau,t) = 100 \times \frac{|N_{\tau,t} \cap H_{\tau,t}|}{|N_{\tau,t}|},$$

$$CLS(\tau,t) = -\frac{1}{|N_{\tau,t}|}\sum_{i \in N_{\tau,t}} log_2 q_i(t \mid \tau).$$

Higher recall indicates better prioritization of the scored buildings, and lower CLS indicates greater geometric-mean normalized probability mass at their locations. Unreported buildings remain in the normalization without being assigned unaffected outcomes. The normalization treats relative affected probability as a proxy for report-location probability, without explicitly modeling spatial differences in reporting propensity. CLS therefore evaluates relative spatial allocation rather than calibration of absolute building-level affected probabilities.

Two 30-min windows are evaluated at 09:00 and 09:30. Figure 6 indicates 6, 32, and 18 reports at the 08:30, 09:00, and 09:30 updates, respectively. For building-level evaluation, repeated reports and buildings reported in an earlier window are counted only once, yielding 31 and 16 newly reported unique buildings in the two windows. The evaluation probabilities are taken from the corresponding open-loop and closed-loop results at each window endpoint.

For 08:30–09:00, open-loop WUI-CAST places 9 of the 31 subsequently reported buildings in the highest-risk 10%, whereas closed-loop WUI-CAST places 26 in that set. Recall therefore increases from 29.03% to 83.87%, a gain of 54.84 percentage points. CLS decreases from 12.4456 to 11.7474 bits, a reduction of 0.6982 bits. For 09:00–09:30, recall increases from 0 of 16 to 7 of 16 buildings, or from 0.00% to 43.75%, while CLS decreases from 13.6064 to 12.6388 bits, a reduction of 0.9676 bits. Table 4 summarizes the forecast-window tests in Panel A and the cumulative-report and block-scale results discussed in Section 4.2.2 in Panels B and C.

The two metrics thus improve in both evaluated windows: assimilation changes the priority ranking and allocates more normalized probability mass to subsequently reported locations. This extends the evidence beyond agreement with buildings already assimilated. Newly reported buildings need not have become affected during the forecast interval, because discovery and reporting can lag physical fire involvement.

**Table 4.** Paired open-loop and closed-loop performance of WUI-CAST.

| Metric | Evaluation times or set | Open-loop | Closed-loop PF | Change (closed-loop − open-loop) |
|---|---|---|---|---|
| **A Building-level performance against newly reported buildings[1]** | | | | |
| Recall@10% ↑ | 08:30 → 09:00; 32 reports; 31 unique newly reported buildings | 29.03% (9/31) | 83.87% (26/31) | +54.84 percentage points |
| Recall@10% ↑ | 09:00 → 09:30; 18 reports; 16 unique newly reported buildings | 0.00% (0/16) | 43.75% (7/16) | +43.75 percentage points |
| CLS ↓ (bits) | 08:30 → 09:00; same 31 unique newly reported buildings | 12.4456 | 11.7474 | −0.6982 bits |
| CLS ↓ (bits) | 09:00 → 09:30; same 16 unique newly reported buildings | 13.6064 | 12.6388 | −0.9676 bits |
| **B Building-level performance against cumulative NIST reports** | | | | |
| Recall@10% ↑ | Final time in Figure 6(c) | 13.3% (22/165) | 27.9% (46/165) | +14.5 percentage points |
| CLS ↓ (bits) | 42 equally weighted evaluation times | 13.372 | 12.793 | −0.579 bits |
| **C Community-block spatial agreement** | | | | |
| IoU ↑ | 08:30/09:00/10:00/11:00 PST; (four-time mean) | 0.441/0.652/0.605/0.523 (Mean: 0.555) | 0.439/0.652/0.638/0.537 (Mean: 0.567) | −0.002/0.000/+0.033/+0.014 (Mean: +0.011) |
| Dice ↑ | Same four times; (four-time mean) | 0.613/0.790/0.754/0.687 (Mean: 0.711) | 0.610/0.790/0.779/0.699 (Mean: 0.719) | −0.003/0.000/+0.025/+0.012 (Mean: +0.008) |

Notes: All times are PST. ↑ / ↓ indicate the preferred direction. Changes are calculated from unrounded values as closed-loop minus open-loop; percentage-point changes are not relative percentages. Panel A scores subsequent reports using forecasts evaluated before the next report batch is assimilated. Panels B and C assess agreement with cumulative observations, including previously assimilated reports. Both configurations use the same candidate-building set. CLS is expressed in bits and assesses normalized spatial allocation, not absolute probability calibration. 1. Newly reported refers to report timing, not necessarily the time of physical ignition.

#### 4.2.2 Evolving risk fields and agreement with cumulative reports

The forecast-window results complement the evolving spatial patterns in Figure 6. In the open-loop sequence (Figure 6(a)), elevated risk first concentrates along eastern Paradise and progresses westward and inward. The closed-loop-minus-open-loop maps in Figure 6(b) show how successive report batches redirect this trajectory. Closed-loop updating increases affected-state probabilities in the southwestern, downwind part of the community while reducing probabilities in other areas. This pattern is consistent with correction of a spatial lag in the open-loop trajectory; the maps show redistribution of risk rather than a uniform increase across the domain.

Figure 6(c) evaluates the highest-risk 10% against all unique NIST affected-building reports accumulated by each time. The 09:30 cumulative set, for example, contains 53 buildings: six reported by 08:30, 31 additional buildings by 09:00, and 16 more by 09:30. These cumulative scores include previously assimilated buildings and therefore have a different interpretation from the subsequent-report scores in Section 4.2.1.

Immediately across the 09:00 update, cumulative recall rises from 32.4% to 86.5%. This is a correction following incorporation of new evidence, rather than an open-loop versus closed-loop test on an unseen report set. The active and superseded closed-loop paths show that the benefit changes as the forecast propagates without further updates. At the end of the evaluation period, cumulative recall is approximately 28% for the active closed-loop forecast and 13.3% for open-loop WUI-CAST.

For cumulative CLS, the negative base-2 log of normalized probability is averaged over the cumulative report set at each evaluation time, and the resulting scores are averaged with equal weight across 42 times. The mean decreases from 13.372 bits for open-loop WUI-CAST to 12.793 bits for closed-loop WUI-CAST. The 0.579-bit reduction corresponds to a factor of $2^{0.579} = 1.494$ in the time-weighted geometric-mean normalized probability assigned to reported locations. Because the cumulative sets contain assimilated reports, this result measures spatial agreement with accumulated observations and is not an independent forecast test.

At the block scale, mean IoU increases from 0.555 to 0.567 and mean Dice from 0.711 to 0.719 (Table 4). The closed-loop scores are slightly lower at 08:30, equal at 09:00, and higher at 10:00 and 11:00. The modest mean changes are compatible with larger changes in building-level ranking: aggregation and thresholding convert a continuous probability field into a binary spatial extent and can conceal redistribution within that extent. Spatial overlap, recall, and CLS therefore describe different aspects of the same evolving field.

Taken together, the paired experiments show improved agreement with accumulated reports and improved ranking and spatial probability allocation for subsequent reports in the two tested forecast windows. Section 4.3 examines the model behavior underlying those changes, separating the spatial reach and concentration of posterior updates from their predictive evaluation.

### 4.3 Mechanism of Posterior Updating and Risk Redistribution

Regional and local diagnostics explain how assimilation redistributes risk and uncertainty beyond directly observed buildings; forecasting performance is evaluated in Section 4.2.

#### 4.3.1 Regional state updates and changes in ember exposure

Figure 7(a–c) shows the incremental NIST-derived reports at 08:30, 09:00, and 09:30 PST. To identify buildings whose inferred states respond substantially, we compare open-loop and closed-loop probabilities over the four-state set S = {Safe, Ignite, Fire, Burned}:

$$max_{s \in S} \, |P_i^{CL}(s,t) \, - \, P_i^{OL}(s,t)| \, > \, 0.4.$$

Across the three updates, 56 report records (53 unique reported buildings) are associated with substantial probability changes in more than 500 buildings (Figure 7(d–f)). The adjustment therefore extends beyond the reported locations. At 09:00 and 09:30, many responding buildings align with increases in the scaled ember-exposure field. Within WUI-CAST, updating inferred burning states changes source contributions to ember generation and transport, which in turn changes subsequent exposure. The spatial correspondence illustrates the modeled coupling between updated building states and subsequent ember exposure.

Figure 7(g–l) shows changes in the most-probable building states. At 09:30 PST, a separate regional sample of 396 buildings has a higher mean maximum state probability under closed-loop updating, and the fraction of buildings with a maximum state probability of at least 0.5 increases from 17.7% to 68.2%. More buildings therefore have a dominant inferred state after assimilation.

Posterior dispersion is assessed using the four-state entropy,

$$H_i(t) \, = \, -\sum_{s \in S} P(S_t^i \, = \, s) \, log \, P(S_t^i \, = \, s).$$

Mean four-state entropy decreases by 14.8%, with lower entropy in 89.1% of the regional sample. The posterior becomes more concentrated on average while its spatial organization changes.

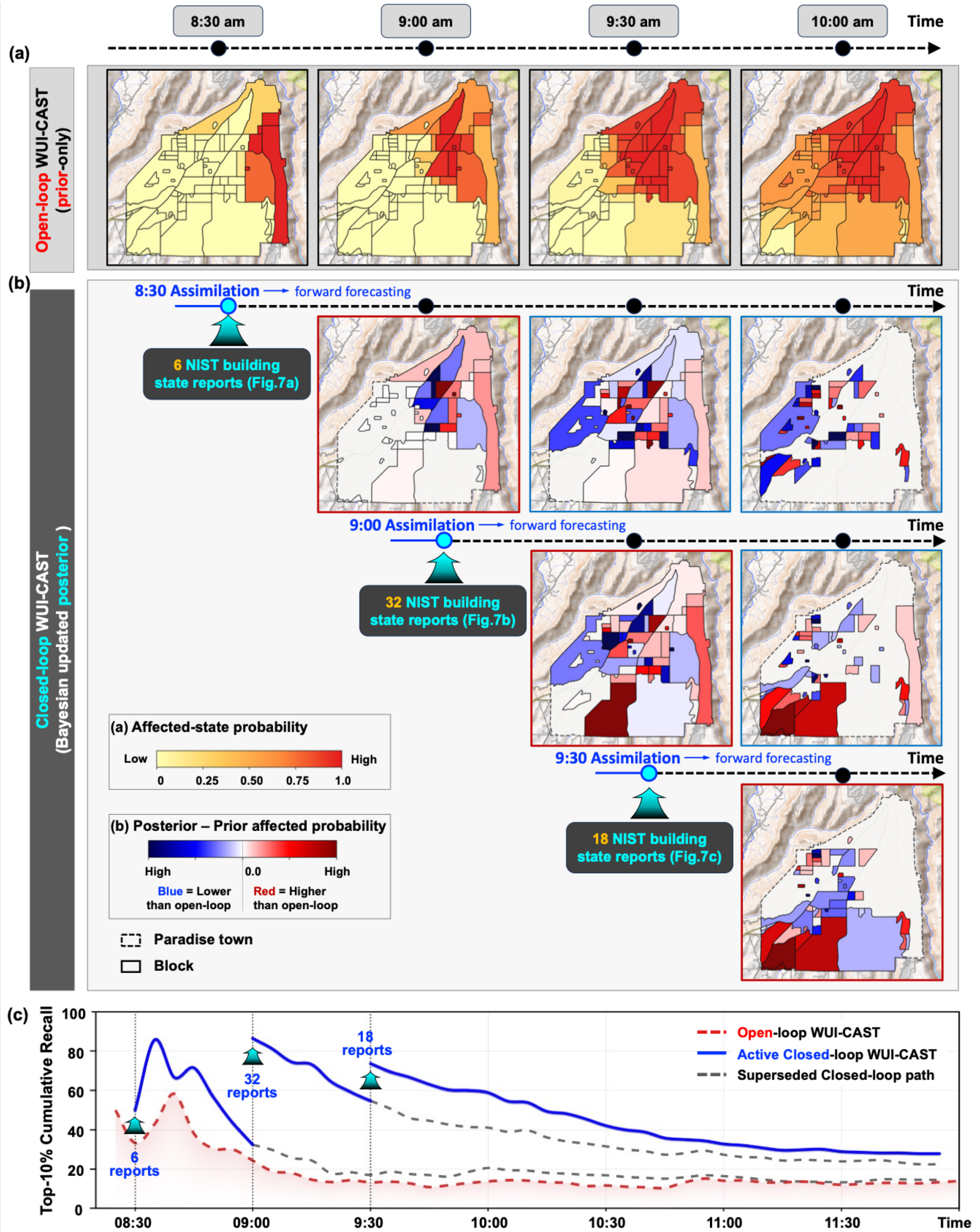

**Figure 6.** Community-scale risk evolution under sequential assimilation. (a) Open-loop hotspot evolution at the block scale. (b) Closed-loop-minus-open-loop changes in affected-state probability along forecast paths initialized by successive NIST-derived report batches; red indicates increased risk relative to open-loop and blue indicates reduced risk. (c) Top-10% cumulative recall for open-loop, active closed-loop, and superseded closed-loop forecast paths. Cumulative recall includes previously assimilated reports; the separate evaluation against subsequent reports is reported in Table 4.

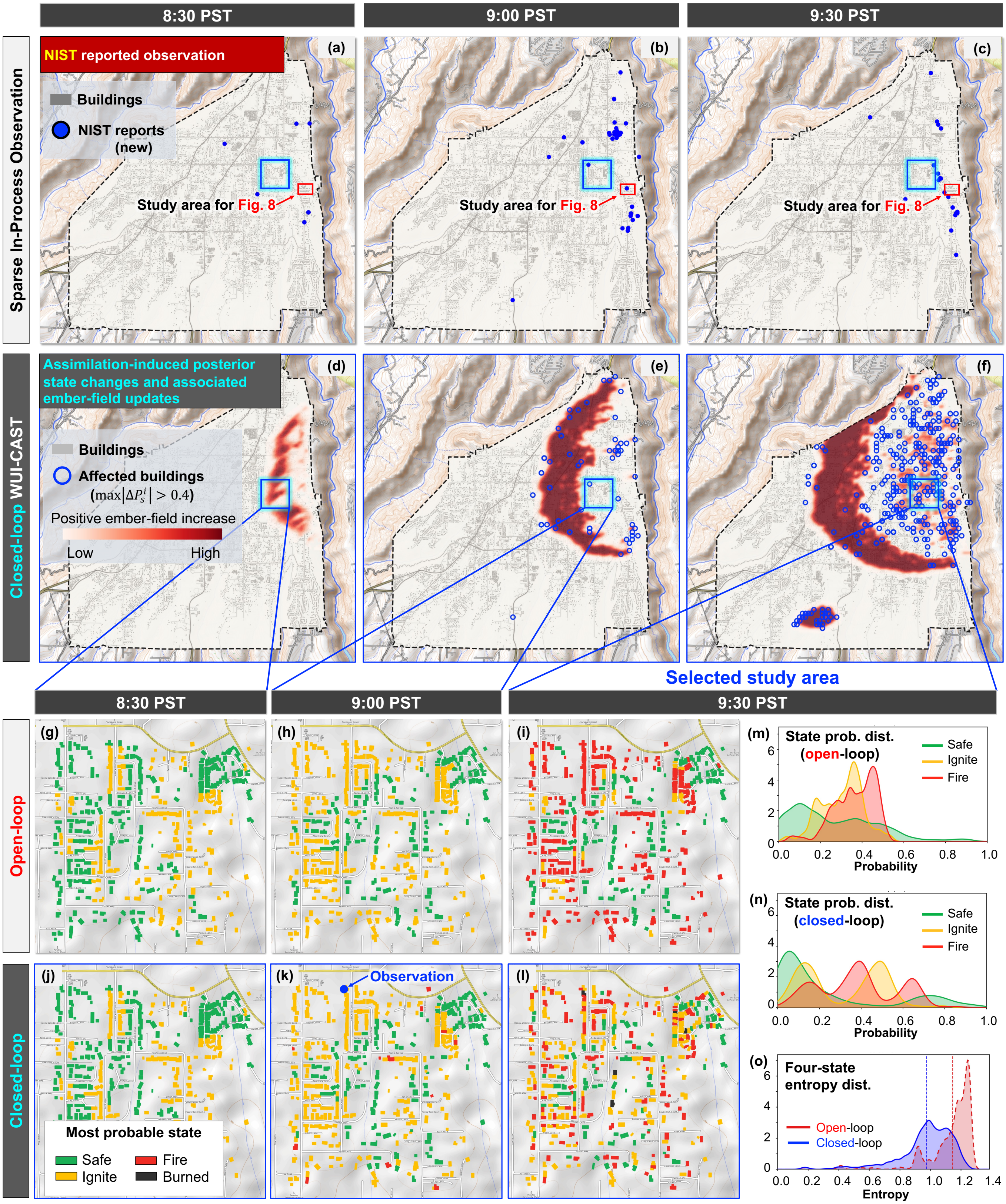

**Figure 7.** Regional posterior response and changes in modeled ember exposure under sparse NIST-derived observations. (a–c) Newly reported observations at 08:30, 09:00, and 09:30 PST. (d–f) Buildings with substantial state-probability changes and positive increases in the scaled ember-exposure field relative to open-loop. (g–l) Open-loop and closed-loop most-probable building states in the selected region. (m–o) Regional statistics at 09:30 PST, including the distributions of state probabilities and four-state entropy.

#### 4.3.2 Local posterior redistribution and building-level state shifts

Figure 8 examines a neighborhood containing one directly observed building, Building A, with a NIST-derived observation at 09:00 PST. By 09:30, the closed-loop solution shows a changed state distribution for Building A and substantial probability differences in 25 nearby buildings under the same 0.4 criterion. This local sample illustrates how posterior changes can extend beyond directly observed structures through the coupled state evolution; its change count is separate from the regional count in Section 4.3.1.

DINS endpoint damage records provide a qualitative reference for this redistribution. Among the 46 neighborhood buildings recorded as burned by the end of the event, assimilation increases both the mean and median affected-state probability at 09:30 and raises the number assigned an affected most-probable state from 13 to 30. The updated field is more compatible with the eventual damage pattern, although endpoint records cannot verify the actual building states at 09:30.

The individual examples in Figure 8(e) show that the response is heterogeneous. For Building A, Fire-state probability increases from 0.45 to 0.90 and entropy decreases from 1.22 to 0.38. Building B shifts from a Safe-dominant open-loop distribution, with P(Safe) = 0.65, to a Fire-dominant closed-loop distribution, with P(Fire) = 0.48. Building C instead shifts toward a safer posterior, with P(Safe) approximately 0.67 and entropy 0.93. For Building D, entropy increases from 0.98 to 1.19 as probability becomes more dispersed across Safe, Ignite, and Fire.

These contrasting responses illustrate selective posterior redistribution: inferred fire pathways may be strengthened in some locations, while risk may decrease elsewhere or uncertainty may increase near ambiguous pathways. Figures 7–8 therefore explain the modeled response accompanying the performance differences in Section 4.2. Section 4.4 examines its sensitivity to observation conditions and filtering implementation.

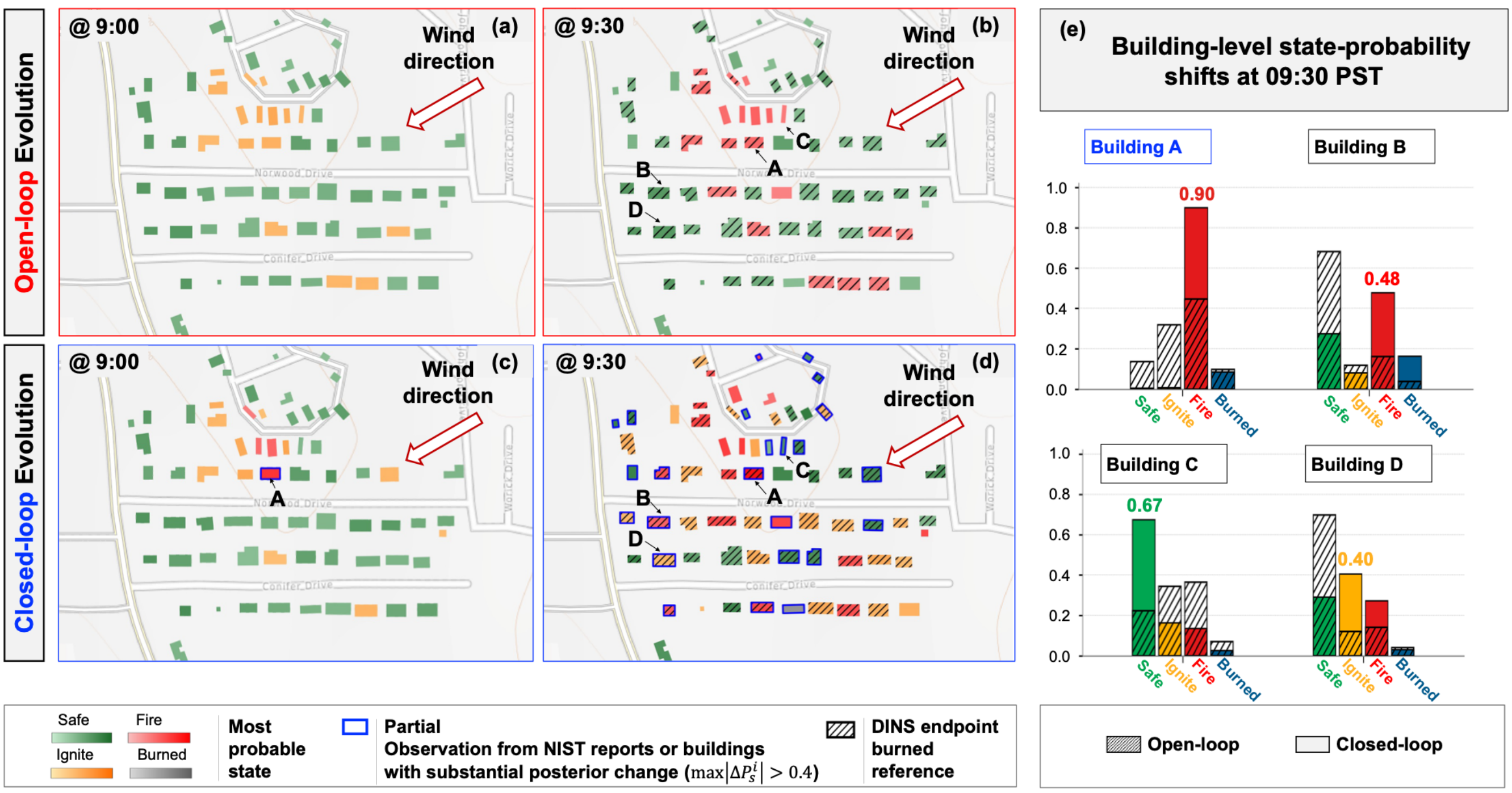

**Figure 8.** Local posterior redistribution in a neighborhood containing one directly observed building. (a–b) Open-loop most-probable states from 09:00 to 09:30 PST. (c–d) Closed-loop most-probable states from 09:00 to 09:30 PST, with Building A directly observed at 09:00 PST. (e) State-probability changes for representative buildings at 09:30 PST. Blue outlines identify substantial posterior changes. DINS endpoint-burned buildings provide a qualitative outcome reference.

### 4.4 Sensitivity to Observation Conditions and Filtering Implementation

The final experiments evaluate the sensitivity of assimilation gains to observation-likelihood specification, observation availability, spatial coverage, and filtering implementation. Figure 9(a) varies the observation likelihood. Figure 9(b) uses a common pseudo-observation pool to vary observation density and spatial coverage while comparing the primary particle filter (PF) with an ensemble Kalman filter (EnKF).

#### 4.4.1 Observation-likelihood sensitivity

The reference likelihood is perturbed along two dimensions: likelihood sharpness and affected-reporting tendency. Across the tested specifications, mean closed-loop cumulative Recall@10% ranges from 49.7% to 56.3%, compared with approximately 18% for the open-loop baseline. Final recall ranges from 27.6% to 32.8%, compared with approximately 13% for open-loop (Figure 9(a)). Sharper likelihoods generally produce stronger corrections, while flatter or higher-reporting specifications produce more conservative changes. The advantage in this cumulative-report metric therefore persists across the selected likelihood settings.

#### 4.4.2 Observation availability and alternative filtering

In a separate controlled sensitivity experiment, the pseudo-observation pool in Figure 9(b-1) combines post-fire building damage status from DINS with coarse arrival-time information inferred from fire-location polygons delineated using NIST reports. The pool provides a common sampling basis for controlled variation in observation quantity and location, but is not treated as independently observed, time-resolved ground truth. Pseudo-observations were sampled from source windows ending at 08:30, 09:00, and 10:00 and then assimilated. PF and EnKF were evaluated at the same prescribed observation-density and spatial-coverage levels. Recall for both methods was calculated against the same complete pool of affected buildings and compared with the same open-loop baseline. Because the evaluation pool includes buildings sampled for assimilation, these experiments assess sensitivity to observation conditions within the constructed pool rather than out-of-sample predictive accuracy against independent, time-resolved observations.

Both filters were evaluated using the same coupled forecast model, 5-min time step, observation schedule, affected-report likelihood, evaluation pool, and prescribed observation-density and spatial-coverage settings. The PF used 5,000 particles, with likelihood-based weight updates and systematic resampling when the effective sample size fell below half the particle count. The EnKF used 5,000 ensemble members; discrete building states were represented as four-component vectors during the update and then mapped back to valid states without backward transitions.

Recall generally increases with observation fraction (Figure 9(b-2)). For PF, the largest improvement occurs between very sparse and moderate observation density, followed by progressively smaller gains. At a fixed 10% observation fraction, restricting observations from unrestricted coverage to fewer accessible 500 m blocks, ultimately a single block, leaves mean recall above the open-loop baseline for both filtering methods in every tested coverage configuration (Figure 9(b-3)).

Across most tested settings, PF has higher mean recall than EnKF. EnKF nevertheless exceeds the open-loop baseline at most tested observation fractions and across all tested spatial-coverage configurations, indicating that assimilation-related improvement is not confined to the primary PF implementation. Differences in plotted uncertainty are not interpreted as intrinsic algorithmic stability because the replication designs differ. The rationale for the primary PF implementation is discussed in Section 5.2.

The following discussion considers the implications of these findings for building-risk forecasting and the limits of their application.

## 5 DISCUSSION

### 5.1 Significance of Closed-Loop Building-Risk Forecasting

The Camp Fire results show that sparse building reports can improve risk prioritization beyond the directly observed locations. The paired open-loop and closed-loop comparisons isolate this benefit within a common propagation model, while the external comparisons provide context for its propagation performance. The practical significance lies in improving which buildings are prioritized as evidence accumulates, even when changes in the aggregate affected extent are modest. This distinction explains why building-level ranking and spatial probability allocation are central to evaluating forecast updating.

The two 30-min tests provide the most direct evidence: both Recall@10% and CLS improve for buildings reported after forecast issuance and scored before their reports are assimilated (Table 4). Thus, the benefit extends beyond agreement with information already incorporated into the model. Cumulative-report scores provide complementary evidence of agreement with accumulated observations. Because reports can lag physical fire involvement, these evaluations concern identification of reported impacts rather than the exact timing of ignition.

The posterior diagnostics help explain this response. Reports increase the weights of trajectories consistent with observed impacts, altering the inferred burning states and the ember exposure that drives subsequent propagation. Risk can therefore increase along some pathways and decrease elsewhere. Local entropy can also increase even when average entropy decreases (Figures 7–8). These heterogeneous responses show that updating redistributes risk and uncertainty rather than simply increasing confidence throughout the community.

### 5.2 Probabilistic Assumptions and the Choice of Assimilation Algorithm

Several probabilistic assumptions were introduced to make recursive inference tractable within the integrated WUI fire model. System evolution was represented as a first-order Markov process, with the next state conditioned on the current system state and external forcing. Conditional on the current state, ember responses, radiative ignition attempts, and observation events across buildings were approximated as conditionally independent.

Ember transport was represented by a parameterized dispersal kernel, deterministic conditional on source states and wind forcing. Embers were assumed to complete transport and deposition within a single simulation time step. A logistic function mapped ember exposure to building ignition probability. Building states evolved irreversibly along the sequence Safe→Ignite→Fire→Burned, with state residence times described by exponential distributions. The associated distributional and transition parameters were derived primarily from previous studies and established empirical understanding.

These approximations enable tractable inference, but the resulting probabilities remain conditional on the transition, ember-transport, and observation models. The sensitivity experiments test selected observation assumptions; they do not establish calibration of absolute building-level probabilities.

PF is the primary implementation because it reweights complete state trajectories using the observation likelihood, accommodating discrete, irreversible, and nonlinear building-state dynamics without imposing a Gaussian posterior. EnKF instead applies approximate corrections based on ensemble-estimated state–observation covariances. The EnKF gains in Section 4.4 demonstrate that the framework supports effective state updating with an alternative filtering implementation.

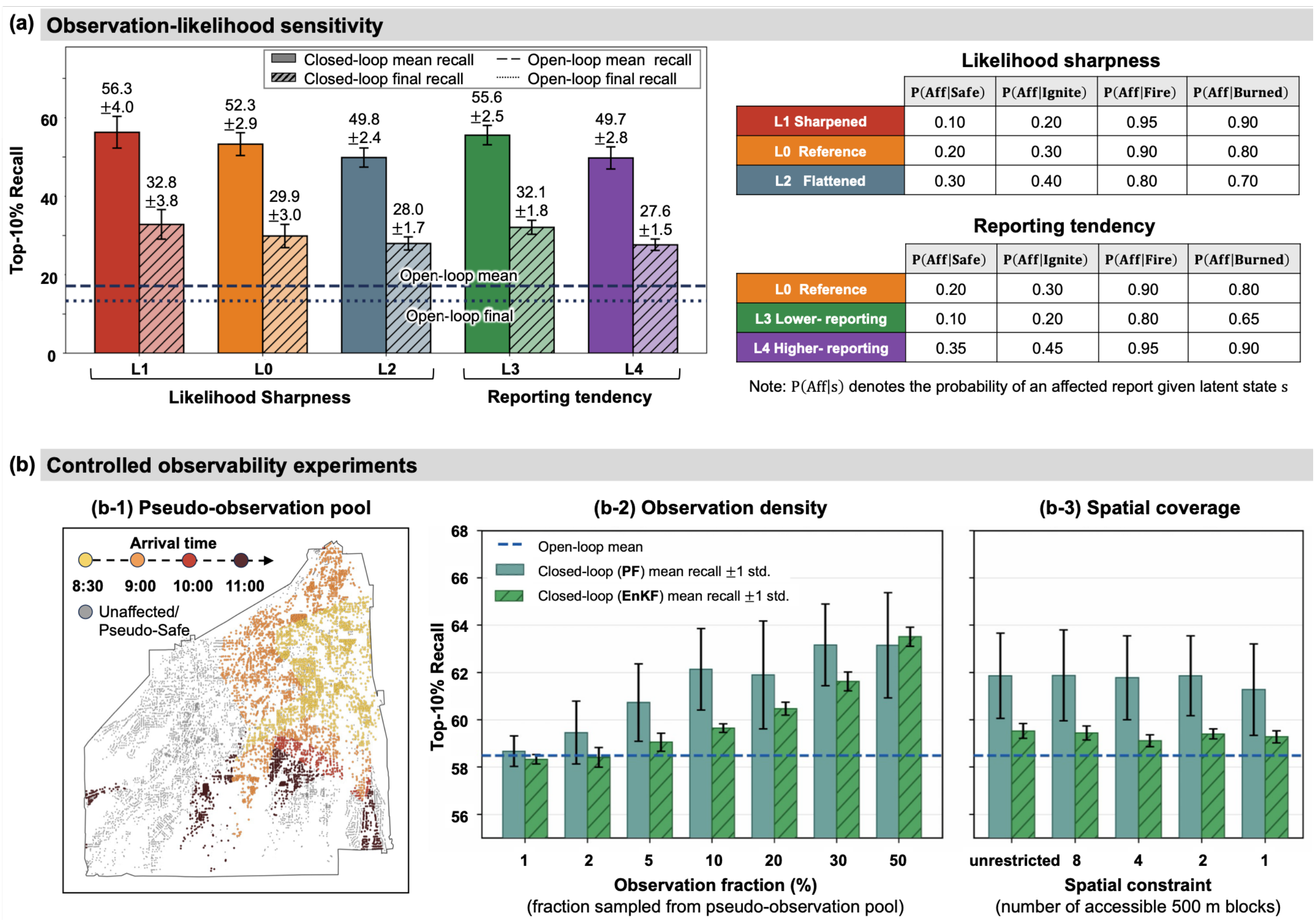

**Figure 9.** Sensitivity to observation-model specification, observation availability, and filtering implementation. (a) Mean and final-step Recall@10% under alternative observation likelihoods; P(Aff | s) is the probability of an affected report conditional on latent state s, and dashed lines denote open-loop baselines. (b-1) Pseudo-observation pool comprising time-indexed affected observations and pseudo-safe locations. (b-2) PF and EnKF performance across observation fractions sampled from that pool. (b-3) Performance under progressively restricted coverage at a fixed 10% observation fraction, with observations confined to fewer accessible 500 m blocks. Bars show mean recall with implementation-specific uncertainty summaries.

### 5.3 Operational Constraints, Model Limitations, and Generalizability

Operational forecasting depends on observation availability and quality. Reports may be delayed, spatially clustered, incorrectly geolocated, duplicated, or inconsistent across sources. The current implementation uses a common observation likelihood and does not explicitly model source-specific reliability, reporting delays, or missing reports. Deployment would require data-quality controls and observation models that distinguish reporting-process errors from changes in the underlying fire state.

The current model also simplifies environmental and structural heterogeneity within the community. Buildings share a common state-transition structure, without explicit differentiation by construction material, structural type, defensible space, or local fuel configuration. Fire suppression, road accessibility, resource deployment, and temporally varying fuel moisture are also excluded from the state evolution. These factors may affect both ignition probability and the observation process, thereby altering the formation and persistence of predicted risk hotspots. Representing building attributes, local environments, and interventions as spatially heterogeneous states or external drivers is an important direction for model development.

On the workstation used in this study (Intel Core Ultra 7 265KF, NVIDIA GeForce RTX 5070 Ti, and 32 GB RAM), the mean computational time per 5-min forecast–update cycle was approximately 42 s for about 1.33 million landscape cells and 11,854 buildings. The mean was calculated over 66 cycles using 100,000 particles and included landscape propagation, coupling between the landscape and building layers, particle prediction and updating, resampling, and building-state probability estimation. The measured cycle time is well below the 5-min update interval, demonstrating computational feasibility for real-time forecast updating at the evaluated scale. Scalability to larger domains and denser communities remains to be evaluated.

The Camp Fire application demonstrates the benefits of sequential updating within a single retrospective event. Evaluation across additional fires will assess the transferability of transition parameters, the suitability of the observation model under different reporting conditions, and the consistency of assimilation-related gains across vegetation, terrain, wind, community structure, and suppression settings. Such comparisons would also distinguish transferable parameters from those requiring local adaptation.

## 6 CONCLUSION

This study presents WUI-CAST, a framework for real-time forecast updating during partially observed wildland–urban interface fires. Its methodological contribution is the integration of established fire-propagation representations with a Dynamic Bayesian Network and sequential Bayesian state estimation. This architecture connects landscape spread, ember-mediated exposure, and building-state evolution, allowing incoming reports to update inferred conditions and reshape subsequent building-risk forecasts.

The retrospective Camp Fire application demonstrates the value of this integration. Open-loop simulations captured key features of the wildfire-to-community transition and achieved community-scale spatial agreement comparable to the published SWUIFT benchmark. Within the same propagation model, sequential assimilation improved identification of subsequently reported buildings. In two 30-min forecasts scored before the evaluation reports were assimilated, Recall@10% increased from 29.03% to 83.87% and from 0.00% to 43.75%, respectively, while conditional spatial logarithmic scores improved in both windows. These results show that assimilation improves both risk prioritization and normalized spatial probability allocation beyond agreement with reports already incorporated into the model.

Regional and local diagnostics explain how sparse observations influence predictions beyond directly observed buildings. Updating the distribution of plausible fire states changes subsequent ember exposure and propagation, increasing inferred risk along some pathways while reducing it elsewhere. Posterior uncertainty also responds selectively, with greater average concentration alongside local increases in ambiguity. This spatial redistribution explains why substantial improvements in building-risk prioritization can accompany relatively modest changes in aggregate affected extent. Controlled experiments further showed assimilation-related gains under alternative observation-likelihood specifications and restricted observation availability, with benefits demonstrated using both particle filtering and EnKF.

A mean computational time of approximately 42 s per 5-min forecast–update cycle on the study workstation supports real-time computational feasibility at the evaluated scale. The Camp Fire proof of concept establishes a foundation for adaptive building-risk forecasting that responds to evolving incident information. Future work will assess transferability across additional fires and incorporate reporting latency, source reliability, and structural heterogeneity, advancing the framework toward operational decision support.

## Acknowledgment

This project is supported by National Fire and Rescue Administration of China (NO. 2025XFZD09).

## REFERENCES

Andrews, P.L., 2007. BehavePlus fire modeling system: Past, present, and future. In: Proceedings of 7th Symposium on Fire and Forest Meteorology; 23-25 October 2007, Bar Harbor, Maine. Boston, MA: American Meteorological Society. 13 p.

Bernard, S., Lebecki, K., Gillard, P., Youinou, L., Baudry, G., 2010. Statistical method for the determination of the ignition energy of dust cloud-experimental validation. Journal of Loss Prevention in the Process Industries 23, 404–411. https://doi.org/10.1016/j.jlp.2010.01.006

Boroujeni, S.P.H., Mehrabi, N., Afghah, F., McGrath, C.P., Bhatkar, D., Biradar, M.A., Razi, A., 2025. Toward AI-driven fire imagery: Attributes, challenges, comparisons, and the promise of VLMs and LLMs. Machine Learning with Applications 22, 100763. https://doi.org/10.1016/j.mlwa.2025.100763

Buechi, H., Weber, P., Heard, S., Cameron, D., Plantinga, A.J., 2021. Long-term trends in wildfire damages in California. International journal of wildland fire 30, 757–762.

Chulahwat, A., Mahmoud, H., Monedero, S., Diez Vizcaíno, F.J., Ramirez, J., Buckley, D., Forradellas, A.C., 2022. Integrated graph measures reveal survival likelihood for buildings in wildfire events. Sci Rep 12, 15954. https://doi.org/10.1038/s41598-022-19875-1

Coen, J.L., Cameron, M., Michalakes, J., Patton, E.G., Riggan, P.J., Yedinak, K.M., 2013. WRF-Fire: Coupled Weather–Wildland Fire Modeling with the Weather Research and Forecasting Model. Journal of Applied Meteorology and Climatology 52, 16–38. https://doi.org/10.1175/JAMC-D-12-023.1

Ding, Y., Saharan, N., Elhami-Khorasani, N., Gernay, T., Zhang, Y., Huang, X., 2026. Coupling Computer-Vision House Fuel Load Estimation with Fire Spread Simulation in the Wildland–Urban Interface. Fire Safety Journal 104713. https://doi.org/10.1016/j.firesaf.2026.104713

Dossi, S., Pastor, E., Àgueda, A., Vacca, P., Planas, E., 2026. Wildfire Natech accidents in the wildland-industrial interface: Exposure pathways analysis. Reliability Engineering & System Safety 275, 112771. https://doi.org/10.1016/j.ress.2026.112771

Esparza, M., Battala, V., Mostafavi, A., 2026. A graph neural network and decision tree modeling approach for predicting wildfire-induced building damage. Computer-Aided Civil and Infrastructure Engineering 100085. https://doi.org/10.1016/j.cacaie.2026.100085

Filkov, A.I., Tihay-Felicelli, V., Masoudvaziri, N., Rush, D., Valencia, A., Wang, Y., Blunck, D.L., Valero, M.M., Kempna, K., Smolka, J., De Beer, J., Campbell-Lochrie, Z., Centeno, F.R., Ibrahim, M.A., Lemmertz, C.K., Tam, W.C., 2023. A review of thermal exposure and fire spread mechanisms in large outdoor fires and the built environment. Fire Safety Journal 140, 103871. https://doi.org/10.1016/j.firesaf.2023.103871

Finney, M.A., 2006. An Overview of FlamMap Fire Modeling Capabilities. In: Andrews, Patricia L.; Butler, Bret W., comps. 2006. Fuels Management-How to Measure Success: Conference Proceedings. 28-30 March 2006; Portland, OR. Proceedings RMRS-P-41. Fort Collins, CO: U.S. Department of Agriculture, Forest Service, Rocky Mountain Research Station. p. 213-220 041.

Finney, M.A., 1998. FARSITE, Fire Area Simulator--model Development and Evaluation. U.S. Department of Agriculture, Forest Service, Rocky Mountain Research Station.

Gneiting, T., Raftery, A.E., 2007. Strictly proper scoring rules, prediction, and estimation. Journal of the American Statistical Association 102, 359–378. https://doi.org/10.1198/016214506000001437

Hamada, M., 1975. Architectural fire resistant themes. Kenchikugaku Taikei, Shokokusha, Tokyo.

Hillen, T., Greese, B., Martin, J., De Vries, G., 2015. Birth-jump processes and application to forest fire spotting. Journal of Biological Dynamics 9, 104–127. https://doi.org/10.1080/17513758.2014.950184

Himoto, K., Akimoto, Y., Hokugo, A., Tanaka, T., 2008. Risk and Behavior of Fire Spread in A Densely-built Urban Area. Fire Saf. Sci. 9, 267–278. https://doi.org/10.3801/IAFSS.FSS.9-267

Himoto, K., Tanaka, T., 2008. Development and validation of a physics-based urban fire spread model. Fire Safety Journal 43, 477–494. https://doi.org/10.1016/j.firesaf.2007.12.008

Jana, D., Malama, S., Szasdi-Bardales, F., Shaik, R.U., Narasimhan, S., Elhami-Khorasani, N., Taciroglu, E., 2025. Improving wildfire resilience of road networks through generative models. Reliability Engineering & System Safety 264, 111429. https://doi.org/10.1016/j.ress.2025.111429

Jiang, W., Wang, F., Fang, L., Zheng, X., Qiao, X., Li, Z., Meng, Q., 2021. Modelling of wildland-urban interface fire spread with the heterogeneous cellular automata model. Environmental Modelling & Software 135, 104895. https://doi.org/10.1016/j.envsoft.2020.104895

Juliano, T.W., Szasdi-Bardales, F., Lareau, N.P., Shamsaei, K., Kosović, B., Elhami-Khorasani, N., James, E.P., Ebrahimian, H., 2024. Brief communication: The Lahaina Fire disaster – how models can be used to understand and predict wildfires. Nat. Hazards Earth Syst. Sci. 24, 47–52. https://doi.org/10.5194/nhess-24-47-2024

Khakzad, N., 2019. Modeling wildfire spread in wildland-industrial interfaces using dynamic Bayesian network. Reliability Engineering & System Safety 189, 165–176. https://doi.org/10.1016/j.ress.2019.04.006

Lareau, N.P., Donohoe, A., Roberts, M., Ebrahimian, H., 2022. Tracking Wildfires With Weather Radars. JGR Atmospheres 127, e2021JD036158. https://doi.org/10.1029/2021JD036158

Lee, S.W., Davidson, R.A., 2010a. Physics-Based Simulation Model of Post-Earthquake Fire Spread. Journal of Earthquake Engineering 14, 670–687. https://doi.org/10.1080/13632460903336928

Lee, S.W., Davidson, R.A., 2010b. Application of a Physics-Based Simulation Model to Examine Post-Earthquake Fire Spread. Journal of Earthquake Engineering 14, 688–705. https://doi.org/10.1080/13632460903336936

Lin, S., Huang, X., Urban, J., McAllister, S., Fernandez-Pello, C., 2019. Piloted ignition of cylindrical wildland fuels under irradiation. Frontiers in Mechanical Engineering 5, 54. https://doi.org/10.3389/fmech.2019.00054

Mahmoud, H., Chulahwat, A., 2020. Assessing wildland–urban interface fire risk. R. Soc. open sci. 7, 201183. https://doi.org/10.1098/rsos.201183

Mahmoud, H., Chulahwat, A., 2018. Unraveling the Complexity of Wildland Urban Interface Fires. Sci Rep 8, 9315. https://doi.org/10.1038/s41598-018-27215-5

Maranghides, A., Link, E., Mell, W. “Ruddy,” Hawks, S., Wilson, M., Brewer, W., Brown, C., Vihnaneck, B., Walton, W.D., 2021. A Case Study of the Camp Fire – Fire Progression Timeline (No. NIST TN 2135). National Institute of Standards and Technology, Gaithersburg, MD. https://doi.org/10.6028/NIST.TN.2135

Maranghides, A., Link, E.D., Mell, W., Hawks, S., Brown, C., Walton, W.D., 2023. A case study of the Camp Fire : notification, evacuation, traffic, and temporary refuge areas (NETTRA) (No. NIST TN 2252). National Institute of Standards and Technology (U.S.), Gaithersburg, MD. https://doi.org/10.6028/NIST.TN.2252

Masoudvaziri, N., Elhami-Khorasani, N., Sun, K., 2023. Toward Probabilistic Risk Assessment of Wildland–Urban Interface Communities for Wildfires. Fire Technol 59, 1379–1403. https://doi.org/10.1007/s10694-023-01382-y

Masoudvaziri, N., Szasdi Bardales, F., Keskin, O.K., Sarreshtehdari, A., Sun, K., Elhami-Khorasani, N., 2021. Streamlined wildland-urban interface fire tracing (SWUIFT): Modeling wildfire spread in communities. Environmental Modelling & Software 143, 105097. https://doi.org/10.1016/j.envsoft.2021.105097

McGrattan, K.B., Baum, H.R., Rehm, R.G., Hamins, A., Forney, G.P., 2000. Fire dynamics simulator- technical reference guide (No. NIST IR 6467). National Institute of Standards and Technology, Gaithersburg, MD. https://doi.org/10.6028/NIST.IR.6467

Mell, W., Maranghides, A., McDermott, R., Manzello, S.L., 2009. Numerical simulation and experiments of burning douglas fir trees. Combustion and Flame 156, 2023–2041. https://doi.org/10.1016/j.combustflame.2009.06.015

MesoWest (n.d.) Weather observations for Jarbo Gap RAWS (JBGC1). Available at https://mesowest.utah.edu/cgi-bin/droman/meso_base_dyn.cgi?stn=JBGC1. Accessed: August 2025.

Microsoft, 2019, March 04. US Building Footprints. https://github.com/microsoft/USBuildingFootprints.

Morandini, F., Perez-Ramirez, Y., Tihay, V., Santoni, P.-A., Barboni, T., 2013. Radiant, convective and heat release characterization of vegetation fire. International Journal of Thermal Sciences 70, 83–91. https://doi.org/10.1016/j.ijthermalsci.2013.03.011

Morvan, D., Hoffman, C., Rego, F., Mell, W., 2011. Numerical simulation of the interaction between two fire fronts in grassland and shrubland. Fire Safety Journal 46, 469–479. https://doi.org/10.1016/j.firesaf.2011.07.008

Ohgai, A., Gohnai, Y., Watanabe, K., 2007. Cellular automata modeling of fire spread in built-up areas—A tool to aid community-based planning for disaster mitigation. Computers, Environment and Urban Systems 31, 441–460. https://doi.org/10.1016/j.compenvurbsys.2006.10.001

Purnomo, D.M.J., Qin, Y., Theodori, M., Zamanialaei, M., Lautenberger, C., Trouvé, A., Gollner, M., 2024. Reconstructing modes of destruction in wildland–urban interface fires using a semi-physical level-set model. Proceedings of the Combustion Institute 40, 105755. https://doi.org/10.1016/j.proci.2024.105755

Rao, K., Konings, A., Yebra, M., Diffenbaugh, N., Williams, P., 2022. The fastest population growth in the West's wildland-urban interface is in areas most vulnerable to wildfires. The Conversation.

Rollins, M.G., 2009. LANDFIRE: a nationally consistent vegetation, wildland fire, and fuel assessment. International Journal of Wildland Fire 18, 235–249. https://doi.org/10.1071/WF08088

Rothermel, R.C., 1972. A Mathematical Model for Predicting Fire Spread in Wildland Fuels. Intermountain Forest & Range Experiment Station, Forest Service, U.S. Department of Agriculture.

Shamsaei, K., Juliano, T.W., Roberts, M., Ebrahimian, H., Kosovic, B., Lareau, N.P., Taciroglu, E., 2023. Coupled fire-atmosphere simulation of the 2018 Camp Fire using WRF-Fire. International Journal of Wildland Fire 32, 195–221. https://doi.org/10.1071/WF22013

Szasdi-Bardales, F., Shamsaei, K., Juliano, T.W., Kosovic, B., Ebrahimian, H., Elhami-Khorasani, N., 2025. An offline coupling of fire spread models to simulate the 2021 Marshall Fire. Int. J. Wildland Fire 34. https://doi.org/10.1071/WF24027

Szasdi-Bardales, F., Shamsaei, K., Lareau, N.P., Juliano, T.W., Kosovic, B., Ebrahimian, H., Elhami-Khorasani, N., 2024. Integrating dynamic wildland fire position input with a community fire spread simulation: A case study of the 2018 Camp Fire. Fire Safety Journal 143, 104076. https://doi.org/10.1016/j.firesaf.2023.104076

Walls, M., Wibbenmeyer, M., 2025. Shaping Land Use Patterns in the Wildland-Urban Interface: The Role of State and Local Governments in Reducing Exposure to Wildfire Risks. Resources for the Future.

Zhao, S., 2011. Simulation of Mass Fire-Spread in Urban Densely Built Areas Based on Irregular Coarse Cellular Automata. Fire Technol 47, 721–749. https://doi.org/10.1007/s10694-010-0187-4